\documentclass[manuscript,screen,review=false,nonacm]{acmart}
\renewcommand\footnotetextcopyrightpermission[1]{} 
\AtBeginDocument{%
  \providecommand\BibTeX{{%
    \normalfont B\kern-0.5em{\scshape i\kern-0.25em b}\kern-0.8em\TeX}}}

\usepackage{tabularx}
\usepackage{subcaption}
\usepackage{tcolorbox}
\usepackage{booktabs}
\usepackage{makecell}
\usepackage{multirow}
\usepackage{array}
\usepackage[normalem]{ulem}
\usepackage{url}
\usepackage{enumitem}
\usepackage{color}
\usepackage{hyperxmp}

\useunder{\uline}{\ul}{}

\newcolumntype{Y}{>{\centering\arraybackslash}X}

\begin{document}

\title{Large Language Models for Cyber Security: A Systematic Literature Review}

\author{Hanxiang Xu}
\affiliation{%
  \institution{Huazhong University of Science and Technology}
  \country{China}
}
\author{Shenao Wang}
\affiliation{%
  \institution{Huazhong University of Science and Technology}
  \country{China}
}
\author{Ningke Li}
\affiliation{%
  \institution{Huazhong University of Science and Technology}
  \country{China}
}
\author{Kailong Wang}
\authornotemark[1]
\affiliation{%
  \institution{Huazhong University of Science and Technology}
  \country{China}
}
\author{Yanjie Zhao}
\affiliation{%
  \institution{Huazhong University of Science and Technology}
  \country{China}
}
\author{Kai Chen}
\authornote{Corresponding authors:~\{wangkl, kchen, haoyuwang\}@hust.edu.cn}
\affiliation{%
  \institution{Huazhong University of Science and Technology}
  \country{China}
}
\author{Ting Yu}
\affiliation{%
  \institution{Hamad Bin Khalifa University}
  \country{The State of Qatar}
}
\author{Yang Liu}
\affiliation{%
  \institution{Nanyang Technological University}
  \country{Singapore}
}
\author{Haoyu Wang}
\authornotemark[1]
\affiliation{%
  \institution{Huazhong University of Science and Technology}
  \country{China}
}
\renewcommand{\shortauthors}{Xu et al.}
\begin{abstract}
The rapid advancement of Large Language Models (LLMs) has opened up new opportunities for leveraging artificial intelligence in a variety of application domains, including cybersecurity. As the volume and sophistication of cyber threats continue to grow, there is an increasing need for intelligent systems that can automatically detect vulnerabilities, analyze malware, and respond to attacks. In this survey, we conduct a comprehensive review of the literature on the application of LLMs in cybersecurity~(LLM4Security). By comprehensively collecting over 40K relevant papers and systematically analyzing 185 papers from top security and software engineering venues, we aim to provide a holistic view of how LLMs are being used to solve diverse problems across the cybersecurity domain.

Through our analysis, we identify several key findings. First, we observe that LLMs are being applied to an expanding range of cybersecurity tasks, including vulnerability detection, malware analysis, and network intrusion detection. Second, we analyze application trends of different LLM architectures (such as encoder-only, encoder-decoder, and decoder-only) across security domains. Third, we identify increasingly sophisticated techniques for adapting LLMs to cybersecurity, such as advanced fine-tuning, prompt engineering, and external augmentation strategies. A significant emerging trend is the use of LLM-based autonomous agents, which represent a paradigm shift from single-task execution to orchestrating complex, multi-step security workflows. Furthermore, we find that the datasets used for training and evaluating LLMs are often limited, highlighting the need for more comprehensive datasets and the use of LLMs for data augmentation. Finally, we discuss the main challenges and opportunities for future research, including the need for more interpretable models, addressing the inherent security risks of LLMs, and their potential for proactive defense. 

Overall, our survey provides a comprehensive overview of the current state-of-the-art in LLM4Security and identifies several promising directions for future research. We believe that the insights and findings presented in this survey will contribute to the growing body of knowledge on the application of LLMs in cybersecurity and provide valuable guidance for researchers and practitioners working in this field. 
\end{abstract}


\maketitle

\section{Introduction}\label{sec:intro}
The rapid advancements in natural language processing (NLP) over the past decade have been largely driven by the development of large language models (LLMs). By leveraging the Transformer architecture~\cite{vaswani2023transformer} and training on massive amounts of textual data, LLMs like BERT~\cite{devlin2019bert}, GPT-3,4~\cite{gpt3.5,gpt4}, PaLM~\cite{chowdhery2022palm}, Claude~\cite{bai2022claude} and Chinchilla ~\cite{hoffmann2022training} have achieved remarkable performance across a wide range of NLP tasks, including language understanding, generation, and reasoning. These foundational models learn rich linguistic representations that can be adapted to downstream applications with minimal fine-tuning, enabling breakthroughs in domains such as open-domain question answering~\cite{abdallah2024generatorretrievergenerator}, dialogue systems~\cite{yi2024survey,ou2024dialogbench}, and program synthesis~\cite{aguinakang2024openuniverse}.

In particular, one important domain where LLMs are beginning to show promise is cybersecurity. With the growing volume and sophistication of cyber threats, there is an urgent need for intelligent systems that can automatically detect vulnerabilities, analyze malware, and respond to attacks~\cite{chen2023vulde4,bilot2023malware3,ids4}. Recent research has explored the application of LLMs across a wide range of cybersecurity tasks, i.e., \textbf{LLM4Security} hereafter. In the domain of software security, LLMs have been used for detecting vulnerabilities from natural language descriptions and source code, as well as generating security-related code, such as patches and exploits. These models have shown high accuracy in identifying vulnerable code snippets and generating effective patches for common types of vulnerabilities~\cite{chow2023vulde6,vulrepair4,charalambous2023vulrepair3}. Beyond code-level analysis, LLMs have also been applied to understand and analyze higher-level security artifacts, such as security policies and privacy policies, helping to classify documents and detect potential violations~\cite{hartvigsen2022harmful5,mets2023harmful4}. In the realm of network security, LLMs have demonstrated the ability to detect and classify various types of attacks from network traffic data, including DDoS attacks, port scanning, and botnet traffic~\cite{ids6,moskal2023cti6,ali2023ids9}. Malware analysis is another key area where LLMs are showing promise, with models being used to classify malware families based on textual analysis reports and behavioral descriptions, as well as detecting malicious domains and URLs~\cite{joyce2023malware1,liu2023ids8}. 

LLMs have also been employed in the field of social engineering to detect and defend against phishing attacks by analyzing email contents and identifying deceptive language patterns~\cite{jamal2023phishing4,phishing6}. Moreover, researchers are exploring the use of LLMs to enhance the robustness and resilience of security systems themselves, by generating adversarial examples for testing the robustness of security classifiers and simulating realistic attack scenarios for training and evaluation purposes~\cite{temara2023pentest3,charan2023pentest4,saha2023hardrepair3}. These diverse applications demonstrate the significant potential of LLMs to improve the efficiency and effectiveness of cybersecurity practices by processing and extracting insights from large amounts of unstructured text, learning patterns from vast datasets, and generating relevant examples for testing and training purposes.

\begin{table}[]
\centering
\caption{State-of-the-art surveys related to LLMs for security.}
\label{tab:intro}
\begin{tabularx}{0.99\linewidth}{l | c |Y | Y | c }
\hline
\textbf{Reference} & \textbf{Year} & \textbf{Scope of topics} & \textbf{Dimensions of discourse} & \textbf{Papers} \\ \hline
Motlagh et al.~\cite{hou2024llm4se1} & 2024 & Security application & Task & Not specified \\ \hline
Divakaran et al~\cite{divakaran2024llm4sec2} & 2024 & Security application & Task & Not specified \\ \hline
Yao et al.~\cite{Yao_2024road3} & 2024 & \makecell[c]{Security application \\ Security of LLM} & \makecell[c]{Model\\ Task} & 281 \\ \hline
Yigit et al.~\cite{yigit2024llm4sec3} & 2024 & \makecell[c]{Security application \\ Security of LLM} & Task & Not specified \\ \hline
Coelho et al.~\cite{dasilva2024securitysurvey} & 2024 & Security application & \makecell[c]{Task\\ Domain specific technique} & 19 \\ \hline
Novelli et al.~\cite{novelli2024llm4sec4} & 2024 & \makecell[c]{Security application \\ Security of LLM} & Task & Not specified \\ \hline
LLM4Security & 2024 & Security application & \makecell[c]{\textbf{Model}\\ \textbf{Task}\\ \textbf{Domain specific technique}\\ \textbf{Data}} & 185 \\ \hline
\end{tabularx}
\end{table}

While there have been several valuable efforts in the literature to survey the LLM4Security~\cite{Yao_2024road3,dasilva2024securitysurvey,motlagh2024llm4sec1,divakaran2024llm4sec2}, given the growing body of work in this direction, these studies often have a more focused scope. Many of the existing surveys primarily concentrate on reviewing the types of tasks that LLMs can be applied to, without providing an extensive analysis of other essential aspects related to these tasks, such as the data and domain-specific techniques employed~\cite{yigit2024llm4sec3,novelli2024llm4sec4}, as shown in Table~\ref{tab:intro}.  For example, Divakaran et al.~\cite{divakaran2024llm4sec2} only analyzed the prospects and challenges of LLMs in various security tasks, discussing the characteristics of each task separately. However, it lacks insight into the connection between the requirements of these security tasks and data, as well as the application of LLMs in domain-specific technologies.

\textbf{Scope.} To address these limitations and provide an in-depth understanding of the state-of-the-art in LLM4Security, we conduct a systematic and extensive survey of the literature. By comprehensively collecting 47,135 relevant papers and systematically analyzing 185 papers from top security and software engineering venues, our survey aims to provide a holistic view of how LLMs are being applied to solve diverse problems across the cybersecurity domain. In addition to identifying the types of tasks that LLMs are being used for, we also examine the specific datasets, preprocessing techniques, and domain adaptation methods employed in each case. This enables us to provide a more nuanced analysis of the strengths and limitations of different approaches, and to identify the most promising directions for future research. Specifically, we focus on answering four key research questions~(RQs):

\begin{itemize}
\item RQ1: What types of security tasks have been facilitated by LLM-based approaches?
\item RQ2: What LLMs have been employed to support security tasks?
\item RQ3: What domain specification techniques are used to adapt LLMs to security tasks?
\item RQ4: What is the difference 
in data collection and pre-processing when applying LLMs to various security tasks?
\end{itemize}

For each research question, we provide a fine-grained analysis of the approaches, datasets, and evaluation methodologies used in the surveyed papers. We identify common themes and categorize the papers along different dimensions to provide a structured overview of the landscape. Furthermore, we highlight the key challenges and limitations of current approaches to guide future research towards addressing the gaps. 

\textbf{Significance.}~A key significance of this survey lies in its deep-seated connection to Software Engineering (SE) community. This connection is not incidental; our systematic selection process deliberately includes premier SE venues (e.g., ICSE, FSE, ASE, etc.), and our analysis reveals that the vast majority of research (63\%) falls under the umbrella of software and system security. The surveyed applications of LLMs are fundamentally addressing core challenges within the secure software development lifecycle, ranging from vulnerability detection and automated program repair to fuzzing and secure code generation. More profoundly, we observe that the methodologies employed in LLM4Security represent an evolution of established SE research paradigms~\cite{hou2024llm4se1}. For instance, the generative capabilities of LLMs are pushing the boundaries of classic SE tasks like automated testing and program repair, while the reliance on curated code corpora and the novel use of LLMs for data augmentation align directly with the principles of data-driven software engineering. Therefore, this survey not only maps a subfield of cybersecurity but also chronicles a significant, ongoing transformation within software engineering itself.

\textbf{Contributions.} The contributions of this work are summarized as follows:
\begin{itemize}
\item We provide the first comprehensive and structured map of the LLM4Security landscape, synthesized from a Systematic Literature Review~(SLR) of 185 high-quality papers selected from an initial pool of over 47,000 candidates.
\item We deliver a multi-dimensional analysis of the field by systematically investigating four crucial aspects across the selected studies: the security tasks being automated, the specific LLM architectures being employed, the domain-specific techniques (e.g., fine-tuning and external augmentation) used for adaptation, and the data-centric practices for training and evaluation.
\item We synthesize and present key insights into how LLMs are reshaping the Secure-Software-Engineering lifecycle. Our findings reveal a paradigm shift from mere defect detection towards automated generation and repair; the use of LLMs for data augmentation to tackle data scarcity; and the emergence of hybrid workflows where LLMs orchestrate traditional SE tools, heralding a new era of security automation.
\end{itemize}

The survey progresses with the following framework. We outline our survey methodology, including the search strategy, inclusion/exclusion criteria, and the data extraction process, in Section~\ref{sec:method}. The analysis and findings for each of the four research questions can be found in Sections~\ref{rq1} through \ref{rq4}. Sections~\ref{sec:threats} to \ref{sec:challenges} explore the constraints and significance of our results, while also identifying promising directions for future research. Finally, Section~\ref{sec:conclusion} concludes the paper.
\section{Methodology}\label{sec:method}
In this study, we conducted a \textbf{Systematic Literature Review (SLR)} to investigate the latest research on \textbf{LLM4Security}. This review aims to provide a comprehensive mapping of the landscape, identifying how LLMs are being deployed to enhance cybersecurity measures. 

\begin{figure}[!htbp]
    \centering
    \includegraphics[width=0.95\textwidth]{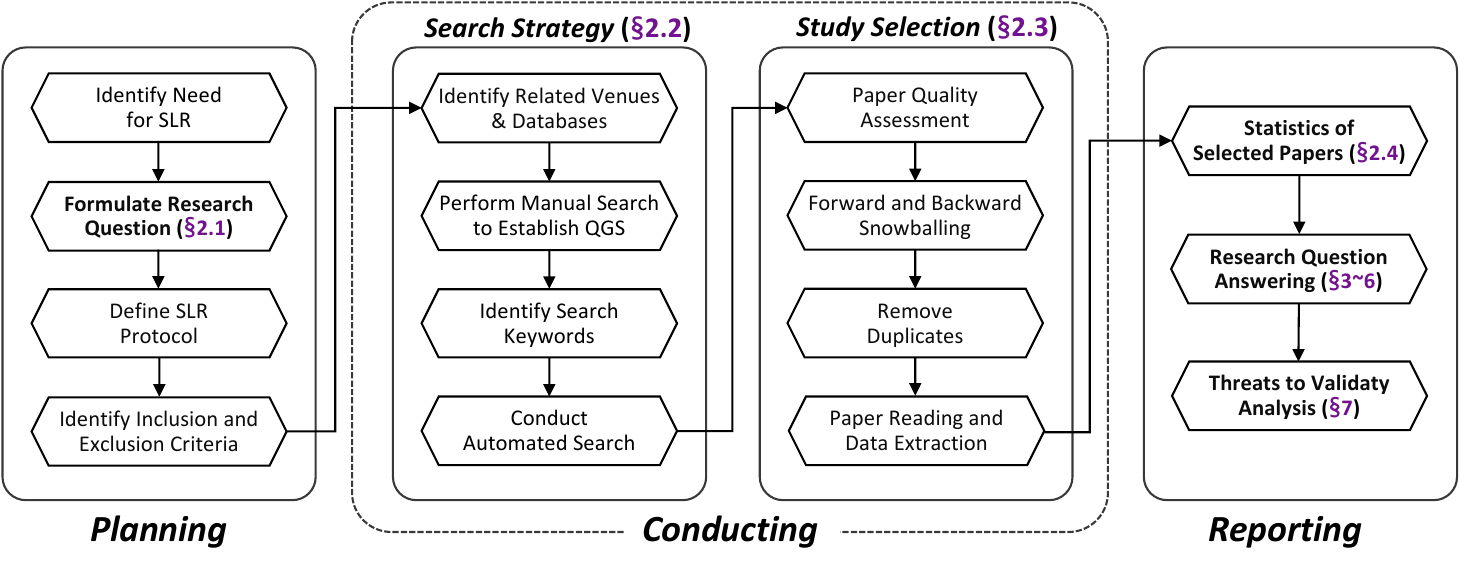}
    \vspace{-5pt}
    \caption{Systematic Literature Review Methodology for LLM4Security.}
    \label{fig:workflow}
\end{figure}

Following the established SLR guidelines~\cite{petersen2015guidelines,kitchenham2009systematic}, our methodology is structured into three pivotal stages as shown in Figure~\ref{fig:workflow}: Planning~(\S\ref{research question}), Conducting~(\S\ref{search strategy}, \S\ref{study selection}), and Reporting~(\S\ref{analysis}), each meticulously designed to ensure comprehensive coverage and insightful analysis of the current state of research in this burgeoning field.

\noindent \textbf{Planning.} Initially, we formulated precise research questions to understand how LLMs are being utilized in security tasks, the benefits derived, and the associated challenges. Subsequently, we developed a detailed protocol delineating our search strategy, including specific venues and databases, keywords, and quality assessment criteria. Each co-author reviewed this protocol to enhance its robustness and align with our research objectives.

\noindent \textbf{Literature survey and analysis.}
We meticulously crafted our literature search to ensure comprehensiveness, employing both manual and automated strategies across various databases to encompass a wide range of papers. Each study identified underwent a stringent screening process, initially based on their titles and abstracts, followed by a thorough review of the full text to ensure conformity with our predefined criteria. To prevent overlooking related papers, we also conducted forward and backward snowballing on the collected papers.

\noindent \textbf{Reporting.} We present our findings through a structured narrative, complemented by visual aids like flowcharts and tables, providing a clear and comprehensive overview of the existing literature. The discussion delves into the implications of our findings, addressing the potential of LLMs to revolutionize cybersecurity practices and identifying gaps that warrant further investigation.

\subsection{Research Question}
\label{research question}
The primary aim of this SLR, focused on the context of LLM4Security, is to meticulously dissect and synthesize existing research at the intersection of these two critical fields. This endeavor seeks to illuminate the multifaceted applications of LLMs in cybersecurity, assess their effectiveness, and delineate the spectrum of methodologies employed across various studies. To further refine this objective, we formulated the following four \textbf{Research Questions (RQs)}:

\begin{itemize}
    \item \textbf{RQ1: What types of security tasks have been facilitated by LLM-based approaches?} Here, the focus is on the scope and nature of security tasks that LLMs have been applied to. The goal is to categorize and understand the breadth of security challenges that LLMs are being used to address, highlighting the model's adaptability and effectiveness across various security dimensions. We will categorize previous studies according to different security domains and provide detailed insights into the diverse security tasks that use LLMs in each security domain.
    
    \item \textbf{RQ2: What LLMs have been employed to support security tasks?} This RQ seeks to inventory the specific LLMs that have been utilized in security tasks. Understanding the variety and characteristics of LLMs used can offer insights into their versatility and suitability for different security applications. We will discuss the architectural differences of LLMs and delve into analyzing the impact of LLMs with different architectures on cybersecurity research over different periods.
    
    \item \textbf{RQ3: What domain specification techniques are used to adapt LLMs to security tasks?} This RQ delves into the specific methodologies and techniques employed to fine-tune or adapt LLMs for security tasks. Understanding these techniques can provide valuable insights into the customization processes that enhance LLMs' effectiveness in specialized tasks. We will elucidate how LLMs are applied to security tasks by analyzing the domain-specific techniques employed in papers, uncovering the inherent and specific connections between these techniques and particular security tasks.
    
    \item \textbf{RQ4: What is the difference in data collection and pre-processing when applying LLMs to security tasks?} This RQ aims to explore the unique challenges and considerations in data processing and model evaluation within the security environment, investigating the correlation between LLMs and the data used for specific tasks. We will reveal the challenges arising from data in applying LLMs to security tasks through two dimensions: data collection and data preprocessing. Additionally, we will summarize the intrinsic relationship among data, security tasks, and LLMs.
\end{itemize}

\subsection{Search Strategy}
\label{search strategy}

To collect and identify a set of relevant literature as accurately as possible, we employed the ``Quasi-Gold Standard'' (QGS) ~\cite{zhang2011identifying} strategy for literature search. The overview of the strategy we applied in this work is as follows:

\begin{figure}[!htbp]
    \centering
    \includegraphics[width=0.95\textwidth]{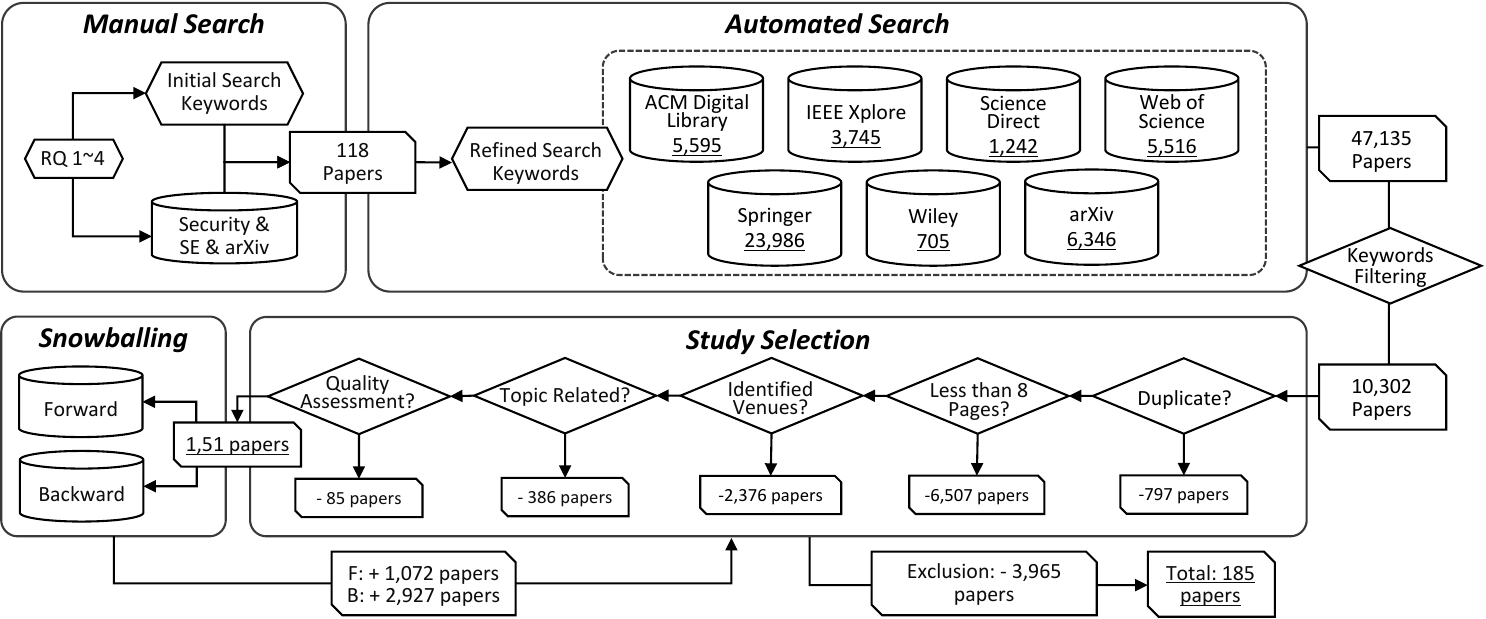}
    \vspace{-5pt}
    \caption{Paper Search and Selection Process.}
    \label{fig:papersearch}
\end{figure}

\noindent \textbf{Step1: Identify related venues and databases.} To initiate this approach, we first identify specific venues for manual search and then choose suitable libraries and databases for the automated search. In this stage, we opt for six of the top Security conferences and journals~(i.e., S\&P, NDSS, USENIX Security, CCS, TDSC, and TIFS) as well as six of the leading Software Engineering conferences and journals~(i.e.,ICSE, ESEC/FSE, ISSTA, ASE, TOSEM, and TSE). Given the emerging nature of LLMs in research, we also include arXiv in both manual and automated searches, enabling us to capture the latest unpublished studies in this rapidly evolving field. For automated searches, we select seven widely utilized databases, namely the ACM Digital Library, IEEE Xplore, Science Direct, Web of Science, Springer, Wiley, and arXiv. These databases offer comprehensive coverage of computer science literature and are commonly employed in systematic reviews within this domain~\cite{hou2024llm4se1,zhan2021research,zhou2024large}.

\noindent \textbf{Step2: Establish QGS.} 
In this step, we start by creating a manually curated set of studies that have been carefully screened to form the QGS. A total of 118 papers relevant to LLM4Sec are manually identified, aligning with the research objective and encompassing various techniques, application domains, and evaluation methods. 

\noindent \textbf{Step3: Define search keywords.} The keywords for automatic search are elicited from the title and abstract of the selected QGS papers through word frequency analysis. The search string consists of two sets of keywords:

\begin{itemize}[leftmargin=15pt]
    \item \textit{Keywords related to LLM: Large Language Model, LLM, Language Model, LM, Pre-trained, CodeX, Llama, GPT-*, ChatGPT, T5, AIGC, AGI.} 
    \item \textit{Keywords related to Security tasks: Cyber Security, Web Security, Network Security, System Security, Software Security, Data Security, Program Analysis, Program Repair, Software Vulnerability, CVE, CWE, Vulnerability Detection, Vulnerability Localization, Vulnerability Classification, Vulnerability Repair, Software Bugs, Bug Detection, Bug Localization, Bug Classification, Bug Report, Bug Repair, Security Operation, Privacy Violation, Denial of Service, Data Poison, Backdoor, Malware Detection, Malware Analysis, Ransomware, Malicious Command, Fuzz Testing, Penetration Testing, Phishing, Fraud, Scam, Forensics, Intrusion Detection.}
\end{itemize}

\noindent \textbf{Step4: Conduct an automated search.} These identified keywords are paired one by one and input into automated searches across the above-mentioned seven widely used databases. Our automated search focused on papers published after 2019, in which GPT-2 was published, as it marked a significant milestone in the development of large language models. The search was conducted in the title, abstract, and keyword fields of the papers in each database. Specifically, the number of papers retrieved from each database after applying the search query and the year filter (2019-2025) is as follows: 5,595 papers in ACM Digital Library, 3,745 papers in IEEE Xplore, 1,242 papers in Science Direct, 5,516 papers in Web of Science, 23,986 papers in Springer, 705 papers in Wiley, and 6,346 papers in arXiv.

\subsection{Study Selection}
\label{study selection}

\begin{table}[!t]
\centering
\caption{Inclusion and exclusion criteria.}
\begin{tabular}{p{0.9\textwidth}}
\hline
\multicolumn{1}{l}{\textbf{Inclusion Criteria}} \\ \hline
\textbf{In\#1:} The title and abstract of the paper contain a pair of identified search keywords; \\
\textbf{In\#2:} Papers that apply large language models~(e.g., BERT, GPT, T5) to security tasks; \\ 
\textbf{In\#3:} Papers that propose new techniques or models for security tasks based on large language models; \\ 
\textbf{In\#4:} Papers that evaluate the performance or effectiveness of large language models in security contexts. \\ \hline
\multicolumn{1}{l}{\textbf{Exclusion Criteria}} \\ \hline
\textbf{Ex\#1:} Duplicate papers, studies with little difference in multi-version from the same authors; \\ 
\textbf{Ex\#2:} Short papers less than 8 pages, tool demos, keynotes, editorials, books, thesis, workshop papers, or poster papers; \\ 
\textbf{Ex\#3:} Papers not published in identified conferences or journals, nor as preprints on arXiv;\\
\textbf{Ex\#4:} Papers that do not focus on security tasks (e.g., natural language processing tasks in general domains); \\ 
\textbf{Ex\#5:} Papers that use traditional machine learning or deep learning techniques without involving large language models; \\ 
\textbf{Ex\#6:} Secondary studies, such as an SLR, review, or survey; \\ 
\textbf{Ex\#7:} Papers not written in English; \\ 
\textbf{Ex\#8:} Papers focus on the security of LLMs rather than using LLMs for security tasks. \\ \hline
\end{tabular}
\label{tab:criteria}
\end{table}

After obtaining the initial pool of 47,135 papers (47,017 from the automated search and 118 from the QGS), we conducted a multi-stage study selection process to identify the most relevant and high-quality papers for our systematic review.

\subsubsection{Coarse-Grained Inclusion and Exclusion Criteria}
To select relevant papers for our research questions, we defined four inclusion criteria and eight exclusion criteria~(as listed in Table~\ref{tab:criteria}) for the coarse-grained paper selection process. Among them, In\#1, Ex\#1, Ex\#2, and Ex\#3 were automatically filtered based on the keywords, duplication status, length, and publication venue of the papers. The remaining inclusion criteria~(In\#2\string~4) and exclusion criteria~(Ex\#4\string~8) were manually applied by inspecting the topic and content of each paper. Specifically, the criteria of In\#1 retained 10,302 papers whose titles and abstracts contained a pair of the identified search keywords. Subsequently, Ex\#1 filtered out 797 duplicate or multi-version papers from the same authors with little difference. Next, the automated filtering criteria Ex\#2 was applied to exclude short papers, tool demos, keynotes, editorials, books, theses, workshop papers, or poster papers, resulting in 6,507 papers being removed. The remaining papers were then screened based on the criteria Ex\#3, which retained 622 full research papers published in the identified venues or as preprints on arXiv. The remaining inclusion and exclusion criteria~(In\#2\string~4, Ex\#4\string~8) were then manually applied to the titles and abstracts of these 622 papers, in order to determine their relevance to the research topic. Three researchers independently applied the inclusion and exclusion criteria to the titles and abstracts. Disagreements were resolved through discussion and consensus. After this manual inspection stage, 236 papers were included for further fine-grained full-text quality assessment.

\subsubsection{Fine-grained Quality Assessment}

To ensure the included papers are of sufficient quality and rigor, we assessed them using a set of quality criteria adapted from existing guidelines for systematic reviews in software engineering. The quality criteria included:

\begin{itemize}[leftmargin=15pt]
    \item \textbf{QAC\#1:} Clarity and appropriateness of research goals and questions;
    \item \textbf{QAC\#2:} Adequacy of methodology and study design;
    \item \textbf{QAC\#3:} Rigor of data collection and analysis processes;
    \item \textbf{QAC\#4:} Validity of results and conclusions;
    \item \textbf{QAC\#5:} Thoroughness of reporting and documentation.
\end{itemize}

Each criterion was scored on a 3-point scale (0: not met, 1: partially met, 2: fully met). Papers with a total score of 6 or higher (out of 10) were considered as having acceptable quality. After the quality assessment, 151 papers remained in the selected set.

\subsubsection{Forward and Backward Snowballing}
\label{snowballing}

To further expand the coverage of relevant literature, we performed forward and backward snowballing on the 151 selected papers. Forward snowballing identified papers that cited the selected papers, while backward snowballing identified papers that were referenced by the selected papers.

Here we obtained 1,072 and 2,927 papers separately during the forward and backward processes. Then we applied the same inclusion/exclusion criteria and quality assessment to the papers found through snowballing. After the initial keyword filtering and deduplication, there were 1,583 papers that remained available. Among them, 77 papers were excluded during the page number filtering step, and 1,472 papers were deleted to ensure the papers were published in the selected venues. After confirming the paper topics and assessing the paper quality, only 34 papers were ultimately retained in the snowballing process, resulting in a final set of 185 papers for data extraction and synthesis.

\subsection{Statistics of Selected Papers}
\begin{figure}[thbp!]
    \centering
    \begin{tabular}{@{\extracolsep{\fill}}c@{}c@{\extracolsep{\fill}}}
            \includegraphics[width=0.5\textwidth]{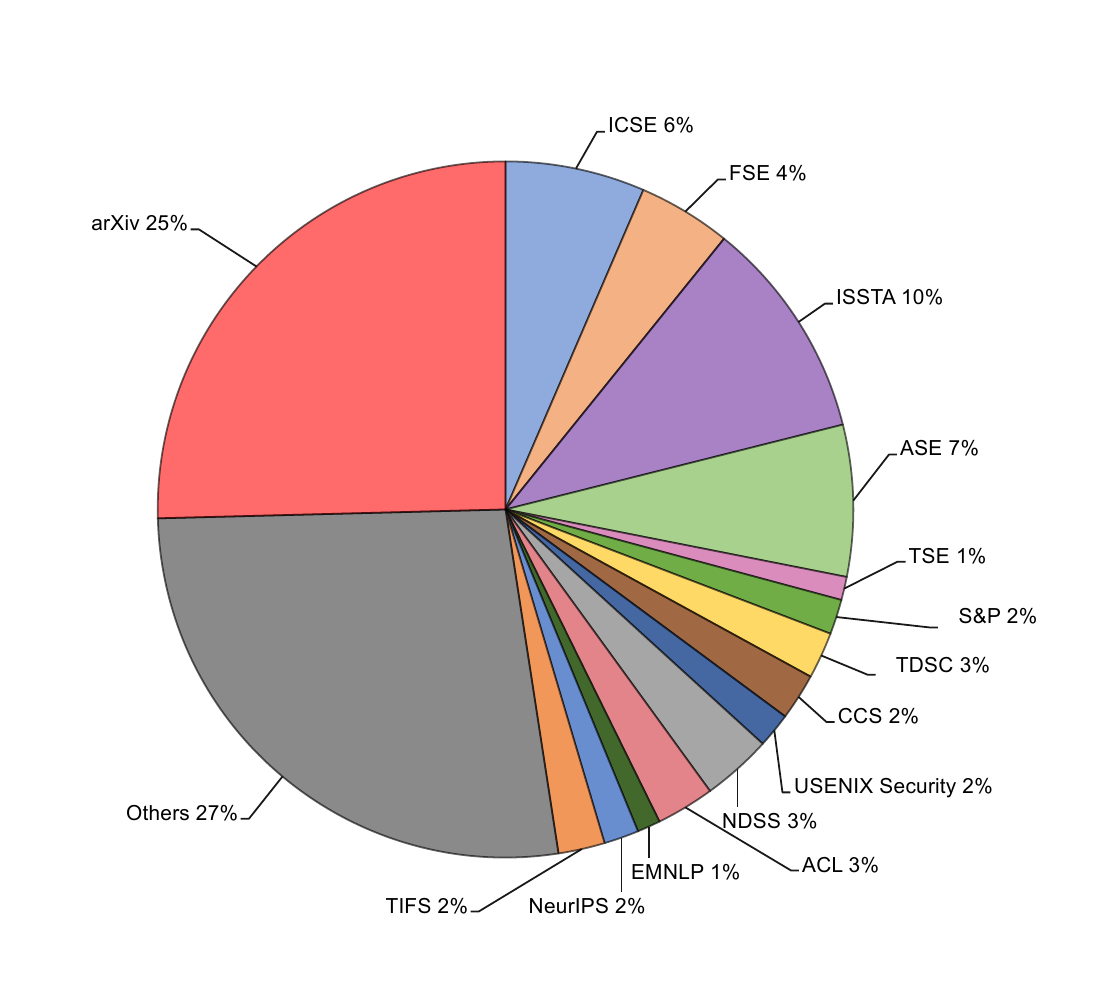} &
            \includegraphics[width=0.5\textwidth]{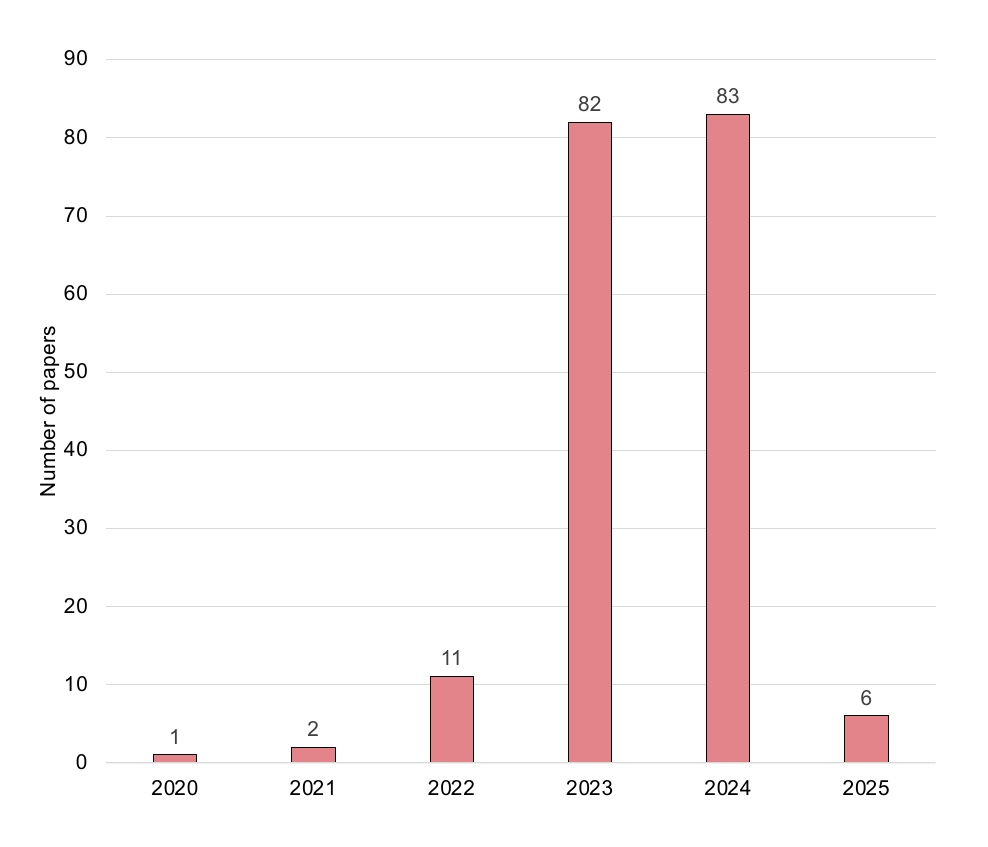}\\
            (A) Distribution of papers across venues. & (B) Distribution of papers over years.\\
    \end{tabular}
    \caption{Overview of the selected 185 papers’ distribution.}
    \label{fig4}
 \end{figure}

After conducting searches and snowballing, a total of 185 relevant research papers were ultimately obtained. The distribution of the included documents is outlined in Figure~\ref{fig4}. As depicted in Figure~\ref{fig4}(A), approximately 75\% of the papers are from peer-reviewed venues or other sources, while the remaining 25\% were published on arXiv, an open-access platform serving as a repository for scholarly articles. Among the specified peer-reviewed venues, ISSTA had the highest frequency (10\%), followed by ICSE (8\%) and ASE (7\%). Other venues making significant contributions included FSE (4\%), NDSS (3\%), ACL (3\%), TDSC (3\%), S\&P (2\%), CCS (2\%), USENIX Security (2\%), TIFS (2\%), NeurIPS (2\%), EMNLP (1\%), and TSE (1\%). The significant portion from arXiv is unsurprising given the rapid emergence of new LLM4Security studies, with many works recently completed and potentially undergoing peer review. Despite lacking peer review in some cases, we conducted rigorous quality assessments on all collected papers to ensure the integrity of our investigation results. This approach enables us to include all high-quality and relevant publications while upholding stringent research standards.

\begin{table}[hbt]
\caption{Extracted data items and related research questions (RQs).}
\begin{tabular}{rl}
\hline
\textbf{RQ} & \textbf{Data Item}                                              \\ \hline
1,2,3,4     & The category of LLM                                             \\
1,3,4       & The category of cybersecurity domain                            \\
1,2,3       & Attributes and suitability of LLMs\\
1,3         & Security task requirements and the application of LLM solutions \\
1           & The security task to which the security domain belongs          \\
3           & Techniques to adapt LLMs to tasks                               \\
3           & Prominent external enhancement techniques                       \\
4           & The types and features of datasets used                         \\
 \hline
\end{tabular}
\label{tab:rq}
\end{table}

The temporal distribution of the included papers is depicted in Figure~\ref{fig4}(B). Since 2020, there has been a notable upward trend in the number of publications. In 2020, only 1 relevant paper was published, followed by 2 in 2021, and 11 papers in 2022. Subsequently, the field experienced explosive growth, with the total count surging to 82 papers in 2023 and a similar peak of 83 papers in 2024. This rapid growth trend signifies an increasing interest in LLM4Security research. Our review also includes 6 early papers from 2025, reflecting the ongoing momentum. We will continue to observe the developments in LLM4Security research.

After completing the full-text review phase, we proceeded with data extraction. The objective was to collect all relevant information essential for offering detailed and insightful answers to the RQs outlined in~\S\ref{research question}. As illustrated in Table~\ref{tab:rq}, the extracted data included the categorization of security tasks, their corresponding domains, as well as classifications of LLMs, external enhancement techniques, and dataset characteristics. Using the gathered data, we systematically examined the relevant aspects of LLM application within the security domains.

\label{analysis}
\section{RQ1: What types of security tasks have been facilitated by LLM-based approaches?}~\label{rq1}

\begin{figure*}[h]
\centering
\includegraphics[width=0.45\textwidth]{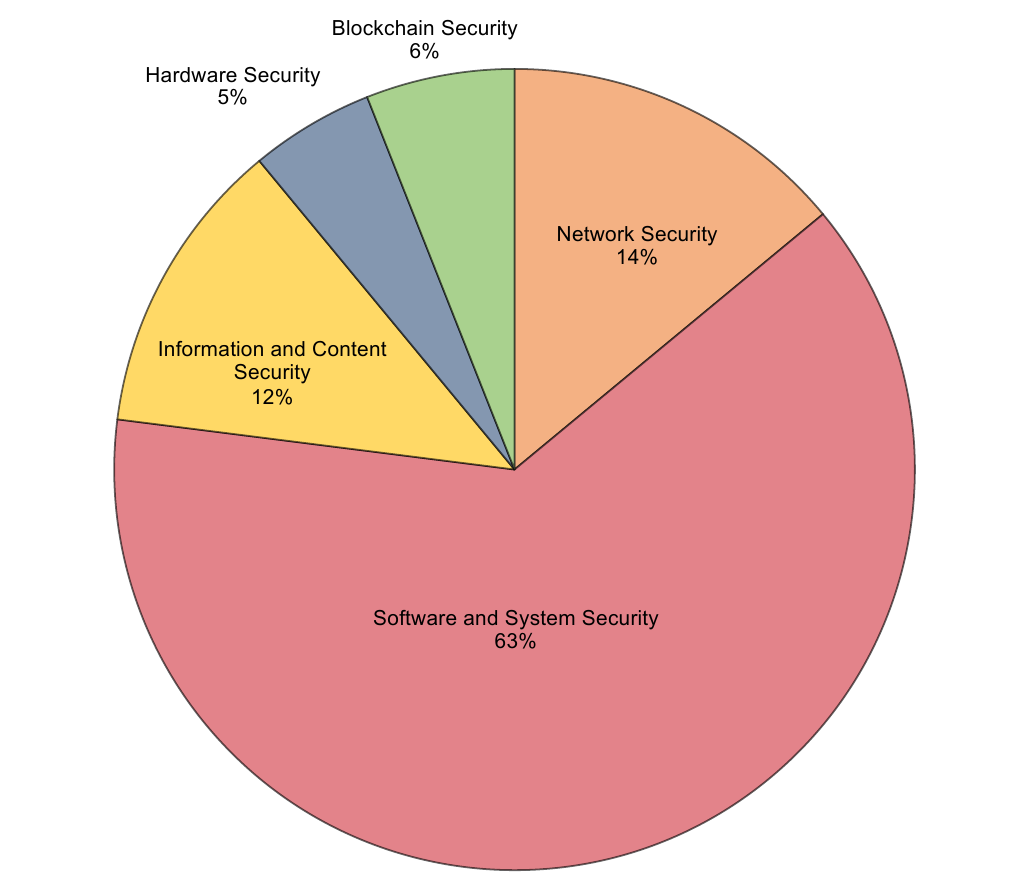}
\caption{Distribution of LLM usages in security domains.}
\label{fig2}
\end{figure*}

This section delves into the detailed examination of LLM utilization across diverse security domains. We have classified them into six primary domains, aligning with the themes of the collected papers: software and system security, network security, information and content security, hardware security, and blockchain security, totaling 185 papers. Figure~\ref{fig2} visually depicts the distribution of LLMs within these six domains. Additionally, Table~\ref{tab2} offers a comprehensive breakdown of research detailing specific security tasks addressed through LLM application. We also analyze the recent and significant trend of employing LLM-based autonomous agents, which represent a paradigm shift from performing isolated tasks to orchestrating complex security workflows.

The majority of research activity in the realm of software and system security, constituting around 63\% of the total research output, is attributed to the advancements made by code LLMs~\cite{sadik2023llm4code1,zheng2024llm4code2,zhang2024llm4code3} and the extensive applications of LLMs in software engineering~\cite{hou2024llm4se1}. This emphasis underscores the significant role and impact of LLMs in software and system security, indicating a predominant focus on leveraging LLMs to automate the handling of potential security issues in programs and systems. Approximately 14\% of the research focus pertains to network security tasks, highlighting the importance of LLMs in aiding traffic detection and network threat analysis. Information and content security activities represent around 12\% of the research output, signaling a growing interest in employing LLMs for generating and detecting fake content. Conversely, activities in hardware security and blockchain security account for approximately 5\% and 6\% of the research output, respectively, suggesting that while exploration in these domains has been comparatively limited thus far, there remains research potential in utilizing LLMs to analyze hardware-level vulnerabilities and potential security risks in blockchain technology.

\begin{table}[htbp]
  \centering
  \caption{Distribution of security tasks over six security domains.}
  \label{tab2}
  \resizebox{0.8\linewidth}{!}{%
  \begin{tabular}{c|l|c} 
  \hline
        \textbf{Security Domains} & \textbf{Security Tasks} & \textbf{Total} \\
        \hline
        \multirow{7}{*}{\textbf{Network Security}} & Web fuzzing~(3) & \multirow{7}{*}{26} \\
                                                  & Traffic and intrusion detection~(10) & \\
                                                  & Cyber threat analysis~(5) & \\
                                                  & Penetration test~(4) & \\
                                                  & Protocol analysis~(1) & \\
                                                  & Protocol fuzzing~(2) & \\
                                                  & Web vulnerability detection~(1) & \\
        \hline
        \multirow{11}{*}{\textbf{Software and System Security}} & Vulnerability detection~(22) & \multirow{11}{*}{119} \\
                                                               & Vulnerability repair~(15) & \\
                                                               & Bug detection~(11) & \\
                                                               & Bug repair~(32) & \\
                                                               & Program fuzzing~(11) & \\
                                                               & Reverse engineering and binary analysis~(10) & \\
                                                               & Malware detection~(3) & \\
                                                               & System log analysis~(10) & \\
                                                               &Secure code generation~(2) & \\
                                                               &ADS security verification~(2) & \\
                                                               &Code obfuscation~(1) & \\
        \hline
        \multirow{5}{*}{\textbf{Information and Content Security}} & Phishing and scam detection~(8) & \multirow{5}{*}{20} \\
                                                                  & Harmful contents detection~(8) & \\
                                                                  & Steganography~(2) & \\
                                                                  & Access control~(1) & \\
                                                                  & Forensics~(1) & \\
        \hline
        \multirow{4}{*}{\textbf{Hardware Security}} & Hardware vulnerability detection~(2) & \multirow{4}{*}{9} \\
                                                  & Hardware vulnerability repair~(5) & \\ 
                                                  & Hardware IP protection~(1) & \\
                                                  & Security assertions generation~(1) & \\
        \hline
        \multirow{2}{*}{\textbf{Blockchain Security}} & Smart contract vulnerability detection~(10) & \multirow{2}{*}{11} \\
                                                    & Transaction anomaly detection~(1) & \\
        \hline
    \end{tabular}
    }
\end{table}

\subsection{Application of LLMs in Network Security}
This section explores the application of LLMs in the field of network security. The tasks include web fuzzing, intrusion and anomaly detection, cyber threat analysis, and penetration testing.

\textbf{Web fuzzing.}~Web fuzzing is a mutation-based fuzzer that generates test cases incrementally based on the coverage feedback it receives from the instrumented web application~\cite{webfuzz3}. Security is undeniably the most critical concern for web applications. Fuzzing can help operators discover more potential security risks in web applications. Liang et al.~\cite{webfuzz2} proposed GPTFuzzer based on a decoder-only architecture. It generates effective payloads for web application firewalls (WAFs) targeting SQL injection, XSS, and RCE attacks by generating fuzz test cases. The model undergoes reinforcement learning~\cite{li2024rewardmodel} fine-tuning and KL-divergence penalty to effectively generate attack payloads and mitigate the local optimum issue. Similarly, Liu et al.~\cite{webfuzz1} utilized an encoder-decoder architecture model to generate SQL injection detection test cases for web applications, enabling the translation of user inputs into new test cases. Meng et al.'s CHATAFL~\cite{meng2024webfuzz4}, on the other hand, shifts focus to leveraging LLMs for generating structured and sequenced effective test inputs for network protocols lacking machine-readable versions. 

\textbf{Traffic and intrusion detection.}~Detecting network traffic and intrusions is a crucial aspect of network security and management~\cite{mirsky2018ids11}. LLMs have been widely applied in network intrusion detection tasks, covering traditional web applications, IoT (Internet of Things), and in-vehicle network scenarios~\cite{ids1,ids2,ids4,ids6}. LLMs not only learn the characteristics of malicious traffic data~\cite{ids4,ids6,ali2023ids9} and capture anomalies in user-initiated behaviors~\cite{ids7} but also describe the intent of intrusions and abnormal behaviors~\cite{ids3,aghaei2023ids5,ali2023ids9}. Additionally, they can provide corresponding security recommendations and response strategies for identified attack types~\cite{chen2023ids10}. Liu et al.~\cite{liu2023ids8} proposed a method for detecting malicious URL behavior by utilizing LLMs to extract hierarchical features of malicious URLs. Their work extends the application of LLMs in intrusion detection tasks to the user level, demonstrating the generality and effectiveness of LLMs in intrusion and anomaly detection tasks.

\textbf{Cyber threat analysis.}~In contemporary risk management strategies, Cyber Threat Intelligence (CTI) reporting plays a pivotal role, as evidenced by recent research~\cite{cti2}. With the continued surge in the volume of CTI reports, there is a growing need for automated tools to facilitate report generation. The application of LLMs in network threat analysis can be categorized into CTI generation and CTI analysis for decision-making. The emphasis on CTI generation varies, including extracting CTI from network security text information (such as books, blogs, news)~\cite{cti1}, generating structured CTI reports from unstructured information~\cite{siracusano2023cti4}, and generating CTI from network security entity graphs~\cite{perrina2023cti5}. Aghaei et al.'s CVEDrill~\cite{aghaei2023cti3} can generate priority recommendation reports for potential cybersecurity threats and predict their impact. Additionally, Moskal et al.~\cite{moskal2023cti6} explored the application of ChatGPT in assisting or automating response decision-making for threat behaviors, demonstrating the potential of LLMs in addressing simple network attack activities.

\textbf{Penetration test.}~Conducting a controlled attack on a computer system to evaluate its security is the essence of penetration testing, which remains a pivotal approach utilized by organizations to bolster their defenses against cyber threats~\cite{schwartz2019pentest2}. The general penetration testing process consists of three steps: information gathering, payload construction, and vulnerability exploitation. Temara~\cite{temara2023pentest3} utilized LLMs to gather information for penetration testing, including the IP address, domain information, vendor technologies, SSL/TLS credentials, and other details of the target website. Sai Charan et al.~\cite{charan2023pentest4} critically examined the capability of LLMs to generate malicious payloads for penetration testing, with results indicating that ChatGPT can generate more targeted and complex payloads for attackers. Happe et al.~\cite{happe2023pentest5} developed an automated Linux privilege escalation guidance tool using LLMs. Additionally, the automated penetration testing tool PentestGPT~\cite{deng2023pentest1}, based on LLMs, achieved excellent performance on a penetration testing benchmark containing 13 scenarios and 182 subtasks by combining three self-interacting modules (inference, generation, and parsing modules).

\textbf{Protocol analysis.}~LLMs are being applied to analyze complex network protocol specifications for detecting inconsistencies. For example, CellularLint utilizes domain-adapted LLMs combined with few-shot and active learning techniques to systematically identify behavioral inconsistencies in 4G/5G cellular standards~\cite{pro_analysis1}. Such inconsistencies found in specifications can impact the security and interoperability of network implementations.

\textbf{Protocol fuzzing.}~LLMs are utilized to guide protocol fuzzing, particularly for protocols specified only in natural language, overcoming limitations of traditional fuzzers in generating structurally and sequentially valid inputs.
Approaches like ChatAFL integrate LLMs into the fuzzing loop, using the model's understanding of message types and sequences to generate valid test inputs and guide mutation-based fuzzers like AFL++~\cite{pro_fuzzing1}. Other frameworks, such as LLMIF, augment LLMs with protocol specifications to automatically extract message formats, field constraints, and state transition rules, and to reason about device responses during fuzzing, specifically targeting IoT protocols~\cite{pro_fuzzing2}. These LLM-guided approaches have demonstrated significant improvements in code coverage and vulnerability discovery compared to baseline fuzzers.

\textbf{Web vulnerability detection.}~LLMs are also employed for detecting vulnerabilities in web applications, although challenges exist regarding language-specific characteristics and data availability. Frameworks like RealVul aim to enhance the performance of LLMs for detecting PHP vulnerabilities~\cite{web_vulde1}. This is achieved by integrating deep program analysis techniques, such as control and data flow analysis, to identify vulnerability candidates and extract relevant code slices.

\subsection{Application of LLMs in Software and System Security}
This section explores the application of LLMs in the field of software and system security. LLMs excel in understanding user commands, inferring program control and data flow, and generating complex data structures~\cite{wu2023llm4sesec}. The tasks it includes vulnerability detection, vulnerability repair, bug detection, bug repair, program fuzzing, reverse engineering and binary analysis, malware detection, and system log analysis.

\textbf{Vulnerability detection.}~The escalation in software vulnerabilities is evident in the recent surge of vulnerability reports documented by Common Vulnerabilities and Exposures (CVEs)~\cite{vulde0}. With this rise, the potential for cybersecurity attacks grows, posing significant economic and social risks. Hence, the detection of vulnerabilities becomes imperative to safeguard software systems and uphold social and economic stability. The method of utilizing LLMs for static vulnerability detection in code shows significant performance improvements compared to traditional approaches based on graph neural networks or matching rules~\cite{chen2023vulde1,ferrag2023vulde2,bakhshandeh2023vulde3,chen2023vulde4,wang2023vulde5,chow2023vulde6,liu-etal-2023vulde7,vulde8,zhang2023vulde9,zhang2023vulde10,khare2023vulde11,ullah2023vulde12,quan2023vulde15}. For instance, SCALE enhances detection by using LLMs to generate comments for Abstract Syntax Tree nodes, creating a novel Structured Natural Language Comment Tree (SCT) representation fed to a graph classifier~\cite{vulde21}. Furthermore, techniques like multi-task instruction fine-tuning, as used in VulLLM, aim to improve model generalization by training simultaneously on detection, localization, and vulnerability explanation tasks~\cite{du-etal-2024-vulde22}. The potential demonstrated by GPT series models in vulnerability detection tasks is particularly evident~\cite{liu-etal-2023vulde7,zhang2023vulde9,khare2023vulde11,chen2023vulde1,ferrag2023vulde2,ullah2024vulde17, vulde20, vulde19, xu2024vulde18}. However, LLMs may generate false positives when dealing with vulnerability detection tasks due to minor changes in function and variable names or modifications to library functions~\cite{ullah2023vulde12}. Liu et al.~\cite{liu2023vulde13} proposed LATTE, which combines LLMs to achieve automated binary taint analysis. This overcomes the limitations of traditional taint analysis, which requires manual customization of taint propagation rules and vulnerability inspection rules. They discovered 37 new vulnerabilities in real firmware. Tihanyi et al.~\cite{tihanyi2023vulde14} used LLMs to generate a large-scale vulnerability-labeled dataset, FormAI, while also noting that over 50\% of the code generated by LLMs may contain vulnerabilities, posing a significant risk to software security.

\textbf{Vulnerability repair.}~Due to the sharp increase in the number of detected vulnerabilities and the complexity of modern software systems, manually fixing security vulnerabilities is extremely time-consuming and labor-intensive for security experts~\cite{zhang2022vulrepair7}. Research shows that 50\% of vulnerabilities have a lifecycle exceeding 438 days~\cite{vulrepair8}. Delayed vulnerability patching may result in ongoing attacks on software systems~\cite{vulrepair9}, causing economic losses to users. The T5 model based on the encoder-decoder architecture performs better in vulnerability repair tasks~\cite{vulrepair4,vulrepair5}. Although LLMs can effectively generate fixes, challenges remain in maintaining the functionality correctness of functions~\cite{vulrepair2}, and they are susceptible to influences from different programming languages~\cite{vulrepair11}. For example, the current capabilities of LLMs in repairing Java vulnerabilities are limited~\cite{vulrepair1}, while studies also investigate their effectiveness on JavaScript vulnerabilities, showing performance improvements with better contextual prompts~\cite{vulrepair11}. Constructing a comprehensive vulnerability repair dataset and fine-tuning LLMs on it can significantly improve the model's performance in vulnerability repair tasks~\cite{vulrepair4, vulrepair14}. To assist developers, especially when vulnerable code locations are untracked in issue reports, approaches like PATUNTRACK use LLMs with Retrieval-Augmented Generation (RAG) and few-shot learning to generate relevant patch examples (insecure code, patch, explanation)~\cite{vulrepair12}. Another direction involves generating natural language repair suggestions instead of direct code patches; VulAdvisor first uses LLMs to create a dataset of suggestions from code/patch pairs and then fine-tunes CodeT5 with context-aware attention to generate such suggestions directly from vulnerable code~\cite{vulrepair13}. Alrashedy et al.~\cite{charalambous2023vulrepair3} proposed an automated vulnerability repair tool driven by feedback from static analysis tools. Tol et al.~\cite{tol2023vulrepair10} proposed a method called ZeroLeak, which utilizes LLMs to repair side-channel vulnerabilities in programs. Charalambous et al.~\cite{alrashedy2024vulrepair6} combined LLMs with Bounded Model Checking (BMC) to verify the effectiveness of repair solutions, addressing the problem of decreased functionality correctness after using LLMs to repair vulnerabilities~\cite{vulrepair15}.

\textbf{Bug detection.}~Bugs typically refer to any small faults or errors present in software or hardware, which may cause programs to malfunction or produce unexpected results. Some bugs may be exploited by attackers to create security vulnerabilities. Therefore, bug detection is crucial for the security of software and system. LLMs can be utilized to generate code lines and compare them with the original code to flag potential bugs within code snippets~\cite{ahmad2023bugde1}. They can also combine feedback from static analysis tools to achieve precise bug localization~\cite{jin2023bugde2, li2023bugde3, bugde9}. Fine-tuning techniques are crucial for bug detection tasks as well, applying fine-tuning allows LLMs to identify errors in code without relying on test cases~\cite{10.1145/bugde5, yang2023bugde8, bugde10}. To address the challenge of hallucinations or false positives in LLM-generated bug reports, frameworks like LLMSAN use techniques such as Chain-of-Thought prompting combined with post-hoc validation of the LLM's claimed evidence (e.g., data-flow paths) via program analysis and targeted queries~\cite{wang-etal-2024-bugde11}. Additionally, Du et al.~\cite{du2023bugde6} and Li et al.~\cite{li2023bugde7} introduced the concept of contrastive learning, which focuses LLMs on the subtle differences between correct and buggy versions of code lines. Fang et al.~\cite{bugde4} proposed a software-agnostic representation method called RepresentThemAll, based on contrastive learning and fine-tuning modules, suitable for various downstream tasks including bug detection and predicting the priority and severity of bugs.

\textbf{Bug repair.}~LLMs possess robust code generation capabilities, and their utilization in engineering for code generation can significantly enhance efficiency. However, code produced by LLMs often carries increased security risks, such as bugs and vulnerabilities~\cite{bugrepair15}. These program bugs can lead to persistent security vulnerabilities. Hence, automating the process of bug fixing is imperative, involving the use of automation technology to analyze flawed code and generate accurate patches to rectify identified issues. LLMs like CodeBERT~\cite{bugrepair5,bugrepair9,huang2023bugrepair10,zhang2023bugrepair11}, CodeT5~\cite{wang2023bugrepair16,tang2023bugrepair14,huang2023bugrepair10,bugrepair25}, Codex~\cite{jin2023bugde2,bugrepair3,xia2023bugrepair13,bugrepair28,bugrepair33}, LLaMa~\cite{tang2023bugrepair14}, CodeLLaMa~\cite{olausson2024bugrepair7,silva2024bugrepair18}
, CodeGEN~\cite{xia2023bugrepair13}, UniXcoder~\cite{zhang2023bugrepair11}, T5~\cite{10.1145/bugrepair4}, PLBART~\cite{huang2023bugrepair10}, Starcoder~\cite{zhao-etal-2024-bugrepair30} and GPT Series~\cite{zhang2023bugrepair2,olausson2024bugrepair7,zhang2023bugrepair11,xia2023bugrepair12,xia2023bugrepair13,tang2023bugrepair14,zhang2023bugrepair17,bugrepair22,bugrepair23,bugrepair24,bugrepair26,bugrepair27,bugrepair29,wang-etal-2024-bugrepair31} have showcased effectiveness in generating syntactically accurate and contextually relevant code. This includes frameworks with encoder-decoder architecture like Repilot~\cite{bugrepair8}, tailored specifically for producing repair patches. Utilizing LLMs for program repair can achieve competitive performance in producing patches for various types of errors and defects~\cite{xia2023bugrepair12}. These models effectively capture the underlying semantics and dependencies in code, resulting in precise and efficient patches. Moreover, fine-tuning LLMs on specific code repair datasets can further improve their ability to generate high-quality patches for real-world software projects. Integrating LLMs into program repair not only speeds up the error-fixing process but also allows software developers to focus on more complex tasks, thereby enhancing the reliability and maintainability of the software~\cite{xia2023bugrepair13}. As demonstrated in the case of ChatGPT, notably enhances the accuracy of program repairs when integrated with interactive feedback loops~\cite{xia2023bugrepair13}. This iterative process of patch generation and validation fosters a nuanced comprehension of software semantics, thereby resulting in more impactful fixes. Complementary to generating fixes, evaluating the capabilities of LLM-driven code agents requires realistic benchmarks; SWT-Bench, for example, provides a framework based on real-world data to jointly assess test generation and fix validation performance~\cite{bugrepair32}. By integrating domain-specific knowledge and technologies with the capabilities of LLMs, their performance is further enhanced. Custom prompts, fine-tuning for specific tasks, retrieving external data, and utilizing static analysis tools~\cite{xia2023bugrepair19,tang2023bugrepair14,vulrepair5,vulrepair4,jin2023bugde2} significantly improve the effectiveness of bug fixes driven by LLMs.

\textbf{Program fuzzing.}~Fuzz testing, or fuzzing, refers to an automated testing method aimed at generating inputs to uncover unforeseen behaviors. Both researchers and practitioners have effectively developed practical fuzzing tools, demonstrating significant success in detecting numerous bugs and vulnerabilities within real-world systems~\cite{progfuzz7}. The generation capability of LLMs enables testing against various input program languages and different features~\cite{xia2024progfuzz4, progfuzz5, progfuzz8}, effectively overcoming the limitations of traditional fuzz testing methods. Under strategies such as repetitive querying, example querying, and iterative querying~\cite{zhang2023progfuzz2}, LLMs can significantly enhance the generation effectiveness of test cases. LLMs can generate test cases that trigger vulnerabilities from historical bug reports of programs~\cite{deng2023progfuzz1}, produce test cases similar but different from sample inputs~\cite{hu2023progfuzz3}, analyze compiler source code to generate programs that trigger specific optimizations~\cite{yang2023progfuzz6}, and split the testing requirements and test case generation using a dual-model interaction framework, assigning them to different LLMs for processing. LLMs are also used to analyze program documentation to predict high-risk option combinations for targeted fuzzing, as demonstrated by ProphetFuzz~\cite{progfuzz9}. Furthermore, LLMs facilitate the automatic generation of fuzz drivers for libraries~\cite{progfuzz10, progfuzz11,xu2024progfuzz12}, for instance, PROMPTFUZZ uses an iterative prompt fuzzing loop guided by code coverage feedback. To improve mutation effectiveness, especially for complex inputs like JavaScript, techniques like CovRL integrate coverage-guided reinforcement learning to fine-tune LLM-based mutators~\cite{progfuzz13}.

\textbf{Reverse engineering and binary analysis.}~Reverse engineering is the process of attempting to understand how existing artifacts work, whether for malicious purposes or defensive purposes, and it holds significant security implications. The capability of LLMs to recognize software functionality and extract important information enables them to perform certain reverse engineering steps~\cite{pearce2022reverse1}. For example, Xu et al.~\cite{xu2023reverse2} achieved recovery of variable names from binary files by propagating LLMs query results through multiple rounds. Techniques like GENNM further advance this by using generative LLMs with context-aware fine-tuning, preference optimization, and iterative inference to recover names from stripped binaries~\cite{reverse8}. Armengol-Estapé et al.~\cite{armengolestapé2024reverse4} combined type inference engines with LLMs to perform disassembly of executable files and generate program source code. LLMs can also be used to assist in binary program analysis. Sun et al.~\cite{reverse3} proposed DexBert for characterizing Android system binary bytecode. Pei et al.~\cite{pei2024reverse5} preserved the semantic symmetry of code based on group theory, resulting in their binary analysis framework SYMC demonstrating outstanding generalization and robustness in various binary analysis tasks. Song et al.~\cite{reverse6} utilized LLMs to address authorship analysis issues in software engineering, effectively applying them to real-world APT malicious software for organization-level verification. Some studies~\cite{Hu2024reverse7, reverse9, reverse10} apply LLMs to enhance the readability and usability of decompiler outputs, including decompiling WebAssembly into higher-level languages like C/C++, thereby assisting reverse engineers in better understanding binary files.

\textbf{Malware detection.}~Due to the rising volume and intricacy of malware, detecting malicious software has emerged as a significant concern. While conventional detection techniques rely on signatures and heuristics, they exhibit limited effectiveness against unknown attacks and are susceptible to evasion through obfuscation techniques~\cite{bilot2023malware3}. LLMs can extract semantic features of malware, leading to more competitive performance. AVScan2Vec, proposed by Joyce et al.~\cite{joyce2023malware1}, transforms antivirus scan reports into vector representations, effectively handling large-scale malware datasets and performing well in tasks such as malware classification, clustering, and nearest neighbor search. To address dataset limitations in specific domains like malicious package detection, approaches like Maltracker combine fine-grained code analysis with LLM-generated synthetic malicious samples to create enhanced datasets for training detection models ~\cite{malware4}. Botacin~\cite{malware2} explored the application of LLMs in malware defense from the perspective of malware generation. While LLMs cannot directly generate complete malware based on simple instructions, they can generate building blocks of malware and successfully construct various malware variants by blending different functionalities and categories. This provides a new perspective for malware detection and defense.

\textbf{System log analysis.}~Analyzing the growing amount of log data generated by software-intensive systems manually is unfeasible due to its sheer volume. Numerous deep learning approaches have been suggested for detecting anomalies in log data. These approaches encounter various challenges, including dealing with high-dimensional and noisy log data, addressing class imbalances, and achieving generalization~\cite{log0}. Nowadays, researchers are utilizing the language understanding capabilities of LLMs to identify and analyze anomalies in log data. Studies compare different strategies, such as supervised fine-tuning (SFT) of models versus few-shot in-context learning (ICL) with LLMs, finding SFT often yields higher accuracy given sufficient labeled data, while ICL is advantageous for rapid deployment or limited data scenarios~\cite{log7}. Compared to traditional deep learning methods, LLMs demonstrate outstanding performance and good interpretability~\cite{qi2023log4,shan2024log6}. Fine-tuning LLMs for specific types of logs~\cite{karlsen2023log3,log8}, using reinforcement learning-based fine-tuning strategies~\cite{han2023log5} can significantly enhance their performance in log analysis tasks. LLMs are also being employed for log analysis in cloud servers~\cite{liu2023log1,chen2023log2}, where their reasoning abilities can be combined with server logs to infer the root causes of cloud service incidents.

\textbf{Secure code generation.}~While LLMs demonstrate strong code generation capabilities, the security of the generated code is a significant concern, as models trained on large corpora may replicate insecure patterns~\cite{bugrepair15}. 
Studies investigating LLM-generated code that utilizes security APIs, such as Java security APIs, have found high rates of misuse, highlighting the risks of relying on LLMs for security-sensitive implementations~\cite{seccode1}. 
To mitigate these risks, research explores methods to guide LLMs towards more secure code generation; for example, PromSec proposes an interactive loop involving static analysis, graph-based code fixing, and reverse-prompt engineering to automatically optimize prompts for generating functionally correct and more secure code~\cite{seccode2}.

\textbf{ADS security verification.}~LLMs are also being applied to the safety and security verification of complex software-intensive systems like Autonomous Driving Systems (ADS). One application involves leveraging LLMs to enhance simulation testing; for instance, SoVAR utilizes LLMs to automatically extract and generalize test scenarios from natural language accident reports, generating more diverse and realistic test cases for simulation~\cite{ads1}. Another application focuses on automating the analysis of test results; DIAVIO employs fine-tuned LLMs, informed by real-world accident data represented in a domain-specific language, to automatically diagnose the cause and type of safety violations identified during ADS simulation testing~\cite{ads2}.

\textbf{Code obfuscation.}~Analyzing obfuscated code is crucial for tasks like malware analysis and understanding protected software, and researchers have begun evaluating LLMs for this purpose. Systematic studies assessing models like the GPT and LLaMA families find that while larger models perform well on non-obfuscated code, their analysis accuracy significantly decreases when dealing with obfuscated code, and they exhibit limited capability in generating de-obfuscated code~\cite{obfuscation1}.


\subsection{Application of LLMs in Information and Content Security}
This section explores the application of LLMs in the field of information and content security. The tasks it includes phishing and scam, harmful contents, steganography, access control, and forensics. 

\textbf{Phishing and scam detection.}~Network deception is a deliberate act of introducing false or misleading content into a network system, threatening the personal privacy and property security of users. Emails, short message service (SMS), and web advertisements are leveraged by attackers to entice users and steer them towards phishing sites, enticing them to click on malicious links~\cite{tang2022phishing0}. LLMs can generate deceptive or false information on a large scale under specific prompts~\cite{phishing6}, making them useful for automated phishing email generation\cite{heiding2023phishing1,roy2024phishing3}, but compared to manual design methods, phishing emails generated by LLMs have lower click-through rates~\cite{heiding2023phishing1}. LLMs can achieve phishing email detection through prompts based on website information~\cite{koide2024phishing2} or fine-tuning for specific email features~\cite{roy2024phishing3,phishing7}. Spam emails often contain a large number of phishing emails. Labonne et al.'s research~\cite{labonne2023phishing5} has demonstrated the effectiveness of LLMs in spam email detection, showing significant advantages over traditional machine learning methods. An interesting study~\cite{cambiaso2023phishing8} suggests that LLMs can mimic real human interactions with scammers in an automated and meaningless manner, thereby wasting scammers' time and resources and alleviating the nuisance of scam emails.

\textbf{Harmful contents detection.}~Social media platforms frequently face criticism for amplifying political polarization and deteriorating public discourse. Users often contribute harmful content that reflects their political beliefs, thereby intensifying contentious and toxic discussions or participating in harmful behavior~\cite{harmful0}. The application of LLMs in detecting harmful content can be divided into three aspects: detection of extreme political stances~\cite{hanley2023harmful1,mets2023harmful4}, tracking of criminal activity discourse~\cite{hu2023harmful2}, and identification of social media bots~\cite{cai2024harmful3, harmful7}, as well as assisting human moderators in rating content against platform policies. For challenging multimodal content like memes, approaches such as EXPLAINHM leverage LLMs to generate explanatory debates between opposing personas to aid in detecting implicit harmfulness~\cite{harmful8}. LLMs tend to express attitudes consistent with the values encoded in the programming when faced with political discourse, indicating the complexity and limitations of LLMs in handling social topics~\cite{hartvigsen2022harmful5}. Hartvigsen et al.~\cite{harmful6} generated a large-scale dataset of harmful and benign discourse targeting 13 minority groups using LLMs. Through validation, it was found that human annotators struggled to distinguish between LLM-generated and human-written discourse, advancing efforts in filtering and combating harmful contents.

\textbf{Steganography.}~Steganography, as discussed in Anderson's work~\cite{anderson1998steganography0}, focuses on embedding confidential data within ordinary information carriers without alerting third parties, thereby safeguarding the secrecy and security of the concealed information. Wang et al.~\cite{steganography1} introduced a method for language steganalysis using LLMs based on few-shot learning principles, aiming to overcome the limited availability of labeled data by incorporating a small set of labeled samples along with auxiliary unlabeled samples to improve the efficiency of language steganalysis. This approach significantly improves the detection capability of existing methods in scenarios with few samples. Bauer et al.~\cite{bauer2022steganography2} used the GPT-2 model to encode ciphertext into natural language cover texts, allowing users to control the observable format of the ciphertext for covert information transmission on public platforms.

\textbf{Access control.}~Access control aims to restrict the actions or operations permissible for a legitimate user of a computer system~\cite{access1}, with passwords serving as the fundamental component for its implementation. Despite the proliferation of alternative technologies, passwords continue to dominate as the preferred authentication mechanism~\cite{10.1007/access3}. PassGPT, a password generation model leveraging LLMs, introduces guided password generation, wherein PassGPT's sampling process generates passwords adhering to user-defined constraints. This approach outperforms existing methods utilizing Generative Adversarial Networks (GANs) by producing a larger set of previously unseen passwords, thereby demonstrating the effectiveness of LLMs in improving existing password strength estimators~\cite{rando2023access2}.

\textbf{Forensics.}~In the realm of digital forensics, the successful prosecution of cybercriminals involving a wide array of digital devices hinges upon its pivotal role. The evidence retrieved through digital forensic investigations must be admissible in a court of law~\cite{selamat2008Forensics2}. Scanlon and colleagues~\cite{Forensics1} delved into the potential application of LLMs within the field of digital forensics. Their exploration encompassed an assessment of LLM performance across various digital forensic scenarios, including file identification, evidence retrieval, and incident response. Their findings led to the conclusion that while LLMs currently lack the capability to function as standalone digital forensic tools, they can nonetheless serve as supplementary aids in select cases.

\subsection{Application of LLMs in Hardware Security}
Modern computing systems are built on System-on-Chip (SoC) architectures because they achieve high levels of integration by using multiple Intellectual Property (IP) cores. However, this also brings about new security challenges, as a vulnerability in one IP core could affect the security of the entire system. While software and firmware patches can address many hardware security vulnerabilities, some vulnerabilities cannot be patched, and extensive security assurances are required during the design process~\cite{10.5555/hardde0}. This section explores the application of LLMs in the field of hardware security. The tasks it includes hardware vulnerability detection and hardware vulnerability repair.

\textbf{Hardware vulnerability detection.}~LLMs can extract security properties from hardware development documents. Meng et al.~\cite{meng2023hardde1} trained HS-BERT on hardware architecture documents such as RISC-V, OpenRISC, and MIPS, and identified 8 security vulnerabilities in the design of the OpenTitan SoC. Additionally, Paria et al.~\cite{paria2023hardde2} used LLMs to identify security vulnerabilities from user-defined SoC specifications, map them to relevant CWEs, generate corresponding assertions, and take security measures by executing security policies.

\textbf{Hardware vulnerability repair.}~LLMs have found application within the integrated System-on-Chip (SoC) security verification paradigm, showcasing potential in addressing diverse hardware-level security tasks such as vulnerability insertion, security assessment, verification, and the development of mitigation strategies~\cite{saha2023hardrepair3}. By leveraging hardware vulnerability information, LLMs offer advice on vulnerability repair strategies, thereby improving the efficiency and accuracy of hardware vulnerability analysis and mitigation efforts~\cite{lin2023hardrepair4}. In their study, Nair and colleagues~\cite{10.1007/hardrepair2} demonstrated that LLMs can generate hardware-level security vulnerabilities during hardware code generation and explored their utility in generating secure hardware code. They successfully produced secure hardware code for 10 CWEs at the hardware design level. Additionally, Tan et al.~\cite{ahmad2023hardrepair1} curated a comprehensive corpus of hardware security vulnerabilities and evaluated the performance of LLMs in automating the repair of hardware vulnerabilities based on this corpus. Further investigation into automated hardware bug repair using LLMs has shown that effective prompt engineering, combined with rigorous functional and formal validation, can successfully fix security bugs in hardware description language code, outperforming synthesis-based tools~\cite{hardrepair6}.

\textbf{Hardware IP protection.}~Beyond detecting and repairing vulnerabilities, the role of LLMs in relation to hardware IP protection is also under investigation, including potential misuse. 
Research such as LLMPirate explores how LLMs might be leveraged for hardware IP piracy, demonstrating automated frameworks that employ techniques like syntax translation and interactive feedback to rewrite hardware netlists for evading detection tools, thus highlighting potential threats and the need for robust defenses~\cite{hardwareip1}.

\textbf{Security assertions generation.}~Assertion-Based Verification (ABV) using languages like SystemVerilog Assertions (SVA) is crucial for hardware security, but manually creating effective security assertions is challenging. 
Research is exploring the use of LLMs to automate this process; studies evaluating LLMs for generating security SVA indicate that while LLMs can produce syntactically correct assertions, the functional correctness and relevance heavily rely on the quality of the prompt context, such as providing related examples or explicit security property descriptions~\cite{assertiongen1}.

\subsection{Application of LLMs in Blockchain Security}
This section explores the application of LLMs in the field of blockchain security. The tasks it includes smart contract security and transaction anomaly detection.

\textbf{Smart contract vulnerability detection.}~With the advancement of blockchain technology, smart contracts have emerged as a pivotal element in blockchain applications~\cite{contrast0}. Despite their significance, the development of smart contracts can introduce vulnerabilities that pose potential risks such as financial losses. While LLMs offer automation for detecting vulnerabilities in smart contracts, the detection outcomes often exhibit a high rate of false positives~\cite{david2023contrast2,chen2023contrast3} and require careful consideration of formal properties~\cite{contract5}. Performance varies across different vulnerability types and is constrained by the contextual length of LLMs~\cite{chen2023contrast3}. GPTLENS~\cite{hu2023contrast4} divides the detection process of smart contract vulnerabilities into two phases: generation and discrimination. During the generation phase, diverse vulnerability responses are generated, and in the discrimination phase, these responses are evaluated and ranked to mitigate false positives. Sun and colleagues~\cite{sun2023contrast1} integrated LLMs and program analysis to identify logical vulnerabilities in smart contracts, breaking down logical vulnerability categories into scenarios and attributes. They utilized LLMs to match potential vulnerabilities and further integrated static confirmation to validate the findings of LLMs. To address risks from code reuse, SOChecker uses LLMs first to complete potentially insecure code snippets from Stack Overflow and then to identify vulnerabilities within the context of the completed code~\cite{contract6}. LLMs are also applied to detect specific complex semantic vulnerabilities, such as Ponzi schemes using zero-shot Chain-of-Thought prompting~\cite{contract7}, or accounting errors through hybrid systems combining LLM-based semantic annotation with rule-based reasoning~\cite{contract8}. Related work also explores using LLMs to generate adversarial contracts for validating exploitability~\cite{contract9} or combining LLMs with GNNs and on-chain analysis to detect deployed malicious contracts before attacks occur~\cite{contract10}.

\textbf{Transaction anomaly detection.}~Due to the limitations of the search space and the significant manual analysis required, real-time intrusion detection systems for blockchain transactions remain challenging. Traditional methods primarily employ reward-based approaches, focusing on identifying and exploiting profitable transactions, or pattern-based techniques relying on custom rules to infer the intent of blockchain transactions and user address behavior~\cite{wu2021blockchainsec2,rodler2018blockchainsec3}. However, these methods may not accurately capture all anomalies. Therefore, more general and adaptable LLMs technology can be applied to effectively identify various abnormal transactions in real-time. Gai et al.~\cite{gai2023blockchainsec1} apply LLMs to dynamically and in real-time detect anomalies in blockchain transactions. Due to its unrestricted search space and independence from predefined rules or patterns, it enables the detection of a wider range of transaction anomalies.

\subsection{Application of LLM Agents in Security}

Beyond single-task applications, a significant recent trend is the development of LLM-based autonomous agents for security. These agents are sophisticated systems built around an LLM core, endowed with capabilities for autonomous reasoning, planning, memory, and the use of external tools to accomplish complex, multi-step goals~\cite{xu2025forewarnedforearmedsurveylarge,luo2025large}. This paradigm shifts from using LLMs as simple query-response tools to employing them as autonomous problem-solvers that can interact with environments and other agents to tackle dynamic security challenges. A predominant architecture emerging is the Multi-Agent System (MAS), where specialized agents collaborate, debate, or operate in a hierarchy to achieve a common objective.

\textbf{Network Security.} In network security, agent-based frameworks are revolutionizing complex offensive tasks. For penetration testing, multi-agent systems like VulnBot~\cite{kong2025vulnbot} and hierarchical teams of agents~\cite{zhu2024teams} automate the entire attack lifecycle. These systems typically feature a planning agent that decomposes the goal, specialized agents that execute tasks like reconnaissance or tool use (e.g., using `nmap` or `Metasploit`), and a reporting agent that synthesizes the findings. This collaborative approach has proven effective in exploiting even zero-day vulnerabilities. For protocol analysis, agents like RFCScan~\cite{zheng2025llmagentfunctionalbug} employ a dual-agent structure, with an "indexing agent" processing RFC documents and a "detection agent" querying this indexed knowledge to find functional bugs.

\textbf{Software and System Security.} This domain has seen a surge in agentic systems that address the full lifecycle of a vulnerability. For vulnerability detection and exploitation, frameworks like FaultLine~\cite{nitin2025faultlineautomatedproofofvulnerabilitygeneration} establish an agentic workflow to automate the generation of proof-of-vulnerability exploits, moving beyond simple detection to active confirmation. For vulnerability repair, multi-agent systems like AutoPatch~\cite{seo2025autopatch} create a pipeline of specialized agents, including a "Vulnerability Verifier" and a "Code Patcher," that work together to automatically fix insecure code generated by LLMs. In proactive defense, SCGAgent~\cite{saul2025scgagentrecreatingbenefitsreasoning} functions as an autonomous agent that leverages security guidelines and generates unit tests to ensure newly written code is both functional and secure. For malware analysis, agents like LAMD~\cite{qian2025lamd} emulate a human analyst's workflow by first using backward program slicing—a classic program analysis technique—to isolate security-critical code regions. This focused context is then fed into a tier-wise reasoning process, allowing the agent to analyze behavior from individual instructions to overall intent and autonomously detect malicious patterns in Android applications. Furthermore, agents are being developed for the complex task of reverse engineering and binary analysis. ReCopilot~\cite{chen2025recopilotreverseengineeringcopilot} is an expert agent trained specifically for binary analysis tasks like function name recovery from decompiled code, outperforming general-purpose models. For end-to-end analysis, Vul-BinLLM~\cite{hussain2025vulbinllmllmpoweredvulnerabilitydetection} detects vulnerabilities directly in compiled binaries by using an optimized decompiler and a function analysis queue to manage the LLM's context window, allowing it to analyze large, real-world binaries while maintaining overall context. A key insight from this research is that the most successful software security agents are not replacing classical tools but orchestrating them; the LLM acts as the intelligent "brain" while requiring the precise "eyes and ears" of static analyzers, debuggers, and program slicers to operate effectively.

\textbf{Information and Content Security.} To combat deception, researchers have developed innovative debate-driven multi-agent systems. In phishing detection, frameworks like PhishDebate~\cite{li2025phishdebatellmbasedmultiagentframework} and similar debate-oriented models~\cite{li2025phishintentionllmuncoveringphishingwebsite} utilize a team of agents. Each agent is assigned a specific role, such as analyzing URL structures, scrutinizing textual content for social engineering tactics, or performing screenshot analysis. These agents then debate their findings to reach a collective, more robust conclusion, significantly reducing false positives and outperforming single-agent approaches. Beyond deception detection, the capabilities of LLM agents also introduce novel challenges and applications. For access control, as agents gain autonomy, the risk of privilege escalation becomes a paramount concern. To address this, systems like Prompt Flow Integrity~\cite{kim2025promptflowintegrityprevent} have been proposed to enforce the principle of least privilege by identifying untrusted data sources and validating unsafe data flows. 

\textbf{Hardware Security.} Hardware security verification has traditionally been a manual, labor-intensive process, making it a prime candidate for automation through LLM agents that can learn the complex semantics of hardware description languages. Frameworks like BugWhisperer~\cite{tarek2025bugwhispererfinetuningllmssoc} are leading examples, applying agent-like automation to identify vulnerabilities in SoC designs. It utilizes a specialized LLM, fine-tuned on a comprehensive hardware vulnerability database, to automatically detect security flaws at the Register-Transfer Level (RTL). Other research directions include RTL-Breaker~\cite{DBLP:conf/date/MankaliBAKMSK25}, which assesses LLM security against backdoor attacks during HDL generation.

\textbf{Blockchain Security.} The immutable and high-value nature of blockchain systems makes smart contract auditing a critical domain where agents show immense potential. The A1 agentic system~\cite{gervais2025ai} exemplifies an execution-grounded approach, transforming a general LLM into an end-to-end exploit generator. The agent hypothesizes a vulnerability, tests an exploit against a forked blockchain state, and uses the deterministic execution feedback to refine its attack, allowing it to overcome hallucinations and discover complex exploits. Other advanced frameworks include Smartify~\cite{karanjai2025multiagentframeworkautomatedvulnerability}, a multi-agent system for detecting and repairing vulnerabilities, and SmartAuditFlow~\cite{wei2025adaptive}, which uses RAG and adaptive planning to improve audit reliability. These systems show that for formal domains like blockchain, the most promising path for agents involves building tight feedback loops between the LLM's hypothesis generation and a deterministic execution harness.

\tcbset{
  title=RQ1 - Summary,  
}

\begin{tcolorbox}
(1) We have divided cybersecurity tasks into six domains: software and system security, network security, information and content security, hardware security, and blockchain security. We have summarized the specific applications of LLMs in these domains.

(2)We discussed 29 cybersecurity tasks and found that LLMs are most widely applied in the field of software and system security, with 119 papers covering 11 tasks.

(3) A significant emerging trend identified is the use of LLM-based autonomous agents. Our analysis highlights that these agents represent a paradigm shift, moving from performing isolated security tasks to orchestrating complex, multi-step workflows across all major security domains.
\end{tcolorbox}

\section{RQ2: What LLMs have been employed to support cybersecurity tasks?}~\label{rq2}

\subsection{Architecture of LLMs in Cybersecurity}
\begin{figure*}[ht!]
\centering
\includegraphics[width=0.99\textwidth]{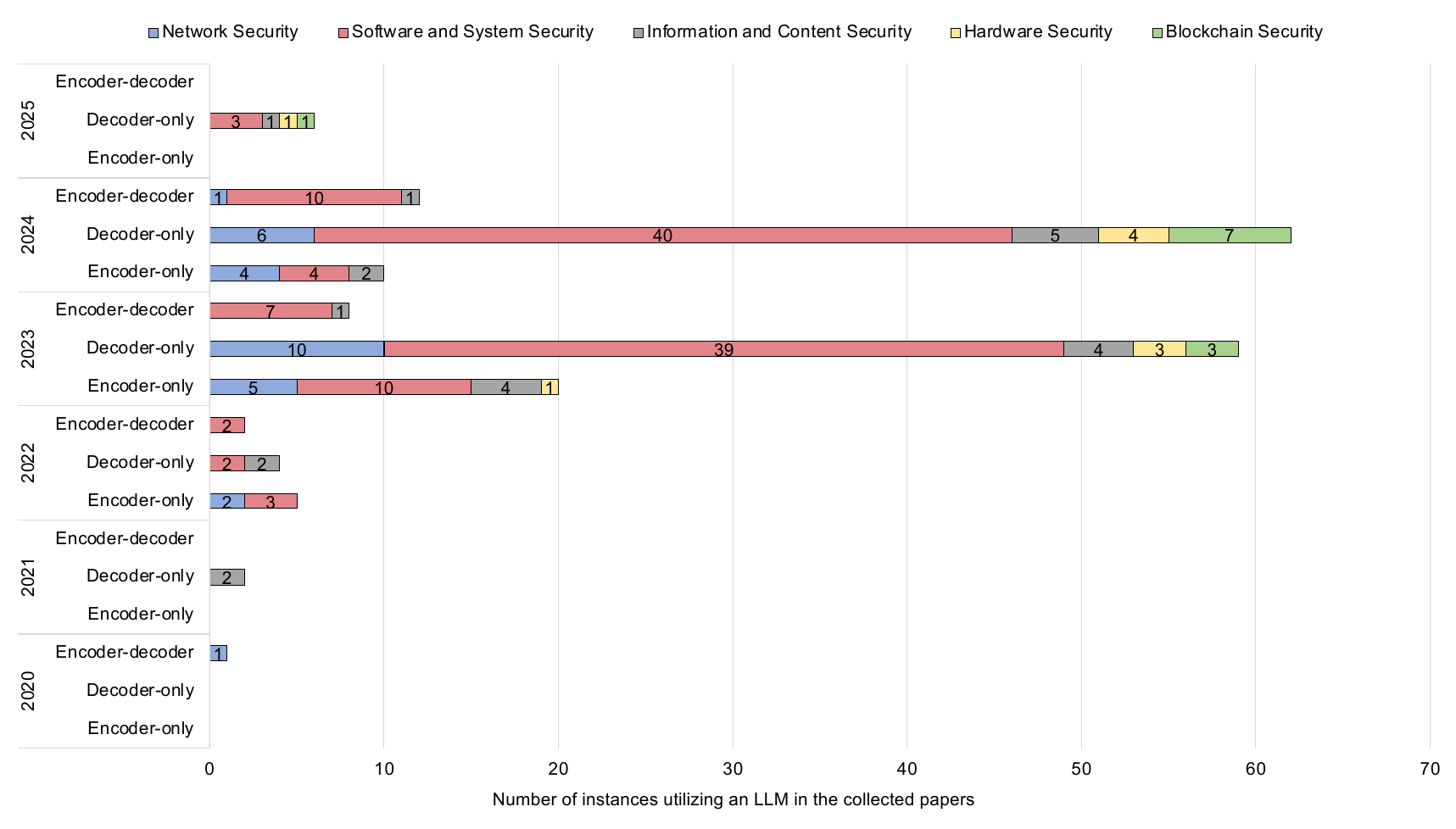}
\caption{Distribution and trend of different model architectures.}
\label{fig:arch_trend}
\end{figure*}

Pre-trained Language Models (PLMs) have exhibited impressive capabilities across various NLP tasks~\cite{shanahan2023talking,kojima2023large,minaee2024large,wei2023chainofthought,zhao2023survey}. Researchers have noted substantial improvements in their performance as model size increases, with surpassing certain parameter thresholds leading to significant performance gains ~\cite{shanahan2023talking,hoffmann2022training}. The term "Large Language Model" distinguishes language models based on the size of their parameters, specifically referring to large-sized PLMs~\cite{minaee2024large,zhao2023survey}.However, there is no formal consensus in the academic community regarding the minimum parameter size for LLMs, as model capacity is intricately linked to training data size and overall computational resources~\cite{kaplan2020scaling}.
In this study, we adopt to the LLM categorization framework introduced by Panet et al.~\cite{Pan_2024}, which classifies the predominant LLMs explored in our research into three architectural categories: encoder-only, encoder-decoder, and decoder-only. We also considered whether the related models are open-source. Open-source models offer higher flexibility and can acquire new knowledge through fine-tuning on specific tasks based on pre-trained models, while closed-source models can be directly called via APIs, reducing hardware expenses. This taxonomy and relevant models are shown in Table~\ref{tab1}. We analyzed the distribution of different LLM architectures applied in various cybersecurity domains, as shown in Fig~\ref{fig:arch_trend}.

\begin{table}[ht]
\caption{The classification of the LLMs used in the collected papers, with the number following the model indicating the count of papers that utilized that particular LLM.}
\centering
\begin{tabularx}{\textwidth}{YYYY}
\toprule 
 & \textbf{Model} & \textbf{Release Time} & \textbf{Open Source} \\
\midrule 
\textbf{Encoder-Only} & BERT~(11) & 2018.10 & Yes \\
                      & RoBERTa~(14) & 2019.07 & Yes \\
                      & DistilBERT~(3) & 2019.10 & Yes\\
                      & CodeBERT~(12) & 2020.02 & Yes\\
                      & DeBERTa~(1)  & 2020.06& Yes \\
                      & GraphCodeBERT~(4) & 2020.09 & Yes\\
                      & CharBERT~(1) & 2020.11 & Yes\\
\midrule
\textbf{Encoder-Decoder} & T5~(4) & 2019.10 & Yes \\
                         & BART~(1) & 2019.10 & Yes \\
                         & PLBART~(4) & 2021.03 & Yes \\ 
                         & CodeT5~(14) & 2021.09 & Yes\\    
                         & UniXcoder~(2) & 2022.03 & Yes\\
                         & Flan-T5~(2) & 2022.10 & Yes \\
\midrule
\textbf{Decoder-Only} & GPT-2~(9) & 2019.02 & Yes \\
                      & GPT-3~(4) & 2020.04 & Yes \\
                      & GPT-Neo~(1) & 2021.03 & Yes \\
                      & CodeX~(14) & 2021.07 & No \\
                      & CodeGen~(6) & 2022.03 & Yes\\
                      & InCoder~(1) & 2022.04 & Yes \\
                      & PaLM~(3) &2022.04 &No\\
                      & Jurassic-1~(1) & 2022.04 & No\\
                      & GPT-3.5~(78) & 2022.11 & No\\
                      & LLaMa~(8) & 2023.02 & Yes \\
                      & GPT-4~(64) & 2023.03 & No \\
                      & Bard~(13) &2023.03 &No \\
                      & Claude~(5) & 2023.03 & No\\
                      & StarCoder~(8) & 2023.05 & Yes\\
                      & Falcon~(2) & 2023.06 & Yes \\
                      & CodeLLaMa~(10) & 2023.08 & Yes\\
                      & Mixtral~(2) & 2023.12 & Yes \\
                      & CodeGemma~(1) & 2024.09 & Yes \\
             
\bottomrule 
\end{tabularx}
\label{tab1}
\end{table}

\textbf{Encoder-only LLMs.}~Encoder-only LLMs, exemplified by BERT~\cite{devlin2019bert} (mentioned in 35 papers in this study) and its variants~\cite{cti1,he2021deberta,liu2019roberta,Ma_2020charbert,feng2020codebert,guo2021graphcodebert,sanh2020distilbert}, primarily focus on understanding and encoding input information rather than generation. In cybersecurity, researchers employ these models primarily to generate contextualized embeddings for relevant data, such as source code, network traffic, or system logs, mapping complex data types into vector representations suitable for downstream tasks~\cite{wang2023vulde5,aghaei2023ids5}.

Various prominent models, including CodeBERT~\cite{feng2020codebert}, GraphCodeBERT~\cite{guo2021graphcodebert}, RoBERTa~\cite{liu2019roberta}, CharBERT~\cite{Ma_2020charbert}, DeBERTa~\cite{he2021deberta}, and DistilBERT~\cite{sanh2020distilbert}, have gained widespread usage due to their ability to effectively process and analyze code and other security-relevant data. CodeBERT~\cite{feng2020codebert}, for instance, is a bimodal model utilizing both natural language and source code inputs, making it suitable for tasks requiring joint understanding, such as vulnerability detection based on code and descriptions. Note that most of these aforementioned BERT variants were not initially designed for cybersecurity tasks; instead, their application in the cybersecurity field stems from their capabilities as general models adapted for code semantics interpretation and understanding. In contrast, SecureBERT~\cite{cti1} is a BERT variant specifically designed for cyber threat analysis tasks, and other domain-specific models like HS-BERT~\cite{meng2023hardde1} have been developed for hardware security property extraction from documentation, demonstrating the adaptation of this architecture for diverse, specialized security needs.

Regarding the model applicability, as shown in the Figure~\ref{fig:arch_trend}, encoder-only models initially garnered attention in the fields of network security~\cite{ids6} and software and systems security~\cite{10.1145/bugde5,bugrepair9}. Specific applications include failure detection in large-scale system logs using BERT fine-tuned on normal behavior~\cite{log8}, and in 2023, this concept was extended to the field of information and content security, utilizing encoder-only models like DistilBERT and RoBERTa for phishing and spam email detection~\cite{jamal2023phishing4} and others to detect harmful content on social media platforms~\cite{mets2023harmful4,hanley2023harmful1,cai2024harmful3}.

\textbf{Encoder-decoder LLMs.}~Encoder-decoder models, such as BART~\cite{lewis2019bart}, T5~\cite{raffel2023T5}, and CodeT5~\cite{wang2021codet5}, structure inputs and outputs, making them suitable for sequence-to-sequence cybersecurity tasks~\cite{vaswani2023transformer}. Specific variants like CodeT5~\cite{wang2021codet5} and PLBART~\cite{ahmad2021plbart} utilize bimodal inputs (programming language and text), enhancing code comprehension for security analysis~\cite{vulrepair1}. This architecture has demonstrated effectiveness in various security applications, including vulnerability repair, where T5-based models like VulRepair~\cite{vulrepair4} achieve strong performance, although effectiveness can vary depending on the language, as noted in studies on Java vulnerability repair using models like CodeT5 and PLBART~\cite{vulrepair1}. Encoder-decoder models like CodeT5 are also utilized within larger frameworks for tasks such as vulnerability detection, for instance, by generating code comments for structural analysis in SCALE~\cite{vulde21}. Furthermore, models like Flan-T5 have been specifically fine-tuned for spam and phishing email detection~\cite{labonne2023phishing5, jamal2023phishing4}. Other applications include program fuzzing~\cite{deng2023progfuzz1} and assisting reverse engineering~\cite{armengolestapé2024reverse4}.

\textbf{Decoder-only LLMs.}~Decoder-only LLMs utilize an autoregressive approach, generating outputs token-by-token, making them well-suited for tasks requiring detailed generation~\cite{radford2018gpt1}, such as security analyses~\cite{log7}, advisories~\cite{obfuscation1}, and code generation~\cite{seccode2}.

Prominent examples include the GPT series (GPT-2~\cite{radford2019gpt2}, GPT-3~\cite{brown2020gpt3}, GPT-3.5~\cite{gpt3.5}, GPT-4~\cite{gpt4}) and specialized versions like Codex~\cite{chen2021codex}. Among these, GPT-3.5 and GPT-4 are the models most frequently used to address various cybersecurity issues in this study, covering almost all cybersecurity applications~\cite{ids2,ids3,Forensics1,deng2023pentest1}. Their strong few-shot learning abilities allow rapid development of new cybersecurity capabilities with minimal fine-tuning. Open-source models like GPT-Neo~\cite{Black2021GPTNeo}, LLaMa~\cite{touvron2023llama}, and Falcon~\cite{penedo2023falcon} also follow this architecture.
Additionally, decoder-only code generation LLMs such as CodeGen~\cite{nijkamp2023codegen}, InCoder~\cite{fried2023incoder}, StarCoder~\cite{li2023starcoder}, Codex~\cite{chen2021codex} and CodeLLaMa~\cite{rozière2024codellama} have been widely used for bug detection and repair, as well as for vulnerability repair~\cite{vulrepair1,xia2023bugrepair12,yang2023bugde8}. The large context window characteristic of decoder-only models allows them to take in and utilize more context relevant to the cybersecurity task, like related vulnerabilities, reports, and code snippets.

Due to the powerful natural language generation capabilities of the decoder-only architecture, researchers initially attempted to apply it to the generation of fake cyber threat intelligence~\cite{phishing6}. Decoder-only LLMs have gained prominence in recent years, especially in 2023 and 2024 as shown in Figure~\ref{fig:arch_trend}, witnessing a surge in development and commercial adoption by leading Internet companies. For instance, Google introduced Bard~\cite{bard}, while Meta unveiled LLaMa~\cite{touvron2023llama}. While GPT-4 and its derivative application, ChatGPT, quickly found integration into various cybersecurity tasks, these other newer models have yet to see similarly widespread adoption in the cybersecurity domain reported in the literature surveyed.

\subsection{Trend Analysis}
Illustrated in Figure~\ref{fig:arch_trend}, from 2020 to 2025 (data available up to early 2025), there have been significant shifts in the preference and utilization of LLM architectures across cybersecurity tasks. The selection of decoder-only, encoder-decoder, and encoder-only structures has influenced diverse research directions and solutions in the cybersecurity field. This examination delves into the trends regarding the adoption of these architectures over time, reflecting the evolving landscape of LLM applications for cybersecurity tasks.

\begin{table}[hbt]
\centering
\caption{Overview of the distribution of LLMs in the open-source community.}
\label{hugtab}
\begin{tabular}{cc}
\begin{subtable}{0.5\textwidth}
\centering
\caption{Top 20 most downloaded models on Huggingface.}
\label{tab:huggingface_downloaded}
\begin{tabular}{cc}
\hline
\multicolumn{1}{c|}{\textbf{Model}}             & \textbf{Architecture} \\ \hline
\multicolumn{1}{c|}{BERT-base}                  & Encoder-only            \\
\multicolumn{1}{c|}{DistilBERT-base}            & Encoder-only            \\
\multicolumn{1}{c|}{GPT2}                       & Decoder-only            \\
\multicolumn{1}{c|}{RoBERTa-large}              & Encoder-only            \\
\multicolumn{1}{c|}{RoBERTa-base}               & Encoder-only            \\
\multicolumn{1}{c|}{xlm-RoBERTa-large}          & Encoder-only            \\
\multicolumn{1}{c|}{xlm-RoBERTa-base}           & Encoder-only            \\
\multicolumn{1}{c|}{DeBERTa-base}               & Encoder-only            \\
\multicolumn{1}{c|}{Qwen-VL-Chat}               & Decoder-only            \\
\multicolumn{1}{c|}{T5-small}                   & Decoder-encoder         \\
\multicolumn{1}{c|}{BERT-base-cased}            & Encoder-only            \\
\multicolumn{1}{c|}{T5-base}                    & Decoder-encoder         \\
\multicolumn{1}{c|}{BERT-base-uncased}          & Encoder-only            \\
\multicolumn{1}{c|}{CamemBERT-base}             & Encoder-only            \\
\multicolumn{1}{c|}{DistilGPT2}                 & Decoder-only            \\
\multicolumn{1}{c|}{DistilRoBERTa-base}         & Encoder-only            \\
\multicolumn{1}{c|}{LLaMa3-8B}                  & Decoder-only            \\
\multicolumn{1}{c|}{ALBERT-base-v2}             & Encoder-only            \\
\multicolumn{1}{c|}{DeBERTa-v3-base}            & Encoder-only            \\
\multicolumn{1}{c|}{ByT5-small}                 & Decoder-encoder         \\ \hline
\end{tabular}
\end{subtable}
\begin{subtable}{0.5\textwidth}
\caption{Top 20 most liked models on Huggingface.}
\label{tab:huggingface_liked}
\centering
\begin{tabular}{cc}
\hline
\multicolumn{1}{c|}{\textbf{Model}}        & \textbf{Architecture} \\ \hline
\multicolumn{1}{c|}{BLOOM-176B}            & Decoder-only             \\
\multicolumn{1}{c|}{LLaMa3-8B}             & Decoder-only             \\
\multicolumn{1}{c|}{LLaMa2-7B}             & Decoder-only             \\
\multicolumn{1}{c|}{Mixtral-8x7B}          & Decoder-only             \\
\multicolumn{1}{c|}{Mixtral-7B}            & Decoder-only             \\
\multicolumn{1}{c|}{Phi-2}                 & Decoder-encoder          \\
\multicolumn{1}{c|}{Gemma-7B}              & Decoder-only             \\
\multicolumn{1}{c|}{ChatGLM-6B}            & Decoder-only             \\
\multicolumn{1}{c|}{StarCoder}             & Decoder-only             \\
\multicolumn{1}{c|}{Falcon-40B}            & Decoder-only             \\
\multicolumn{1}{c|}{Grok-1}                & Decoder-only             \\
\multicolumn{1}{c|}{ChatGLM2-6B}           & Decoder-only             \\
\multicolumn{1}{c|}{GPT2}                  & Decoder-only             \\
\multicolumn{1}{c|}{Dolly-v2-12B}          & Decoder-only             \\
\multicolumn{1}{c|}{BERT-base}             & Encoder-only             \\
\multicolumn{1}{c|}{Zephyr-7B}             & Decoder-only             \\
\multicolumn{1}{c|}{OpenELM}               & Decoder-only             \\
\multicolumn{1}{c|}{Phi-1.5}               & Decoder-encoder          \\
\multicolumn{1}{c|}{Yi-34B}                & Decoder-only             \\
\multicolumn{1}{c|}{Flan-T5}               & Decoder-encoder          \\ 
\hline
\end{tabular}

\end{subtable}
\end{tabular}
\label{hugtab}
\end{table}

In 2020 and 2021, the use of LLMs in cybersecurity was limited, with only 1 and 2 instances identified respectively. In 2020, encoder-decoder LLMs were the sole architecture used (1 instance). However, in 2021, the focus shifted exclusively to decoder-only LLMs (2 instances). This early adoption pattern may be attributed to the research emphasis on LLM performance in natural language processing tasks and innovations in LLM architectures during this period~\cite{kaplan2020scaling,brown2020gpt3}.

The year 2022 marked a significant turning point, with the number of instances employing LLMs for cybersecurity tasks surging to 11, surpassing the combined total from the previous two years. This year also saw increased diversity in the LLM architectures used. Encoder-only LLMs, valued for their representation learning and classification abilities, were utilized in 5 instances (45\%). Encoder-decoder LLMs, with their strong performance on well-defined tasks, were featured in 2 instances (18\%), while decoder-only LLMs, leveraging their knowledge recall and few-shot learning capabilities, garnered 4 instances (36\%). This varied distribution highlights the active exploration of different architectures to address the diverse needs and challenges in cybersecurity.

The years 2023 and 2024 witnessed a dramatic surge in research and a clear shift towards decoder-only LLMs becoming the predominant architecture. This trend is closely tied to the powerful text comprehension, reasoning capabilities~\cite{ouyang2022insgpt,wei2023cot,jiang2020prompt1,dong2023prompt2,madaan2023prompt3}, and open-ended generation demonstrated by chatbots like ChatGPT. These decoder-only models often require minimal fine-tuning and can generate both syntactically correct and functionally relevant code snippets~\cite{laskar2023systematicgpt,sadik2023llm4code1}. In 2023 (total 86 instances), decoder-only models constituted approximately 69\% (59 instances), while encoder-only usage peaked at 23\% (20 instances) and encoder-decoder models represented 8\% (7 instances). This dominance of decoder-only models continued strongly into 2024 (total 85 instances), accounting for roughly 73\% (62 instances). Importantly, unlike the suggestion of complete replacement, both encoder-decoder (12 instances, ~14\%) and encoder-only (11 instances, ~13\%) architectures continued to be utilized in 2024 across various domains, indicating their ongoing relevance for specific tasks. The limited data for early 2025 (6 instances) exclusively features decoder-only models, suggesting their continued prominence.

The dominance of decoder-only LLMs in cybersecurity research aligns with the broader trends in the LLM community. An analysis of the top 20 most liked and downloaded LLMs on Huggingface~\cite{huggingface}, illustrated in Table~\ref{hugtab}), a popular open-source model community, reveals a divergence: encoder-only models like BERT and its variants represent the clear majority of the most downloaded models (13 out of 20), whereas decoder-only models overwhelmingly dominate the most liked list (16 out of 20). This indicates a strong community preference and excitement for the potential of decoder-only models to handle complex, open-ended tasks.
The growing interest in decoder-only LLMs can be attributed to their strong generation, knowledge, and few-shot learning abilities, which make them well-suited for the diverse challenges in cybersecurity. However, the larger parameter size of these models compared to encoder-only models may limit their current adoption due to the scarcity of computational resources~\cite{fedus2022challenge2}.

In summary, while the cybersecurity field leverages all three main LLM architectures, recent years (2023-2024) have shown a clear dominance of decoder-only models, likely driven by their generative strengths and few-shot capabilities popularized by models like ChatGPT. However, the continued application of encoder-only and encoder-decoder models underscores their specific advantages for certain tasks, indicating a diverse rather than monolithic architectural landscape in LLM4Security.

\tcbset{
  title=RQ2 - Summary,  
}

\begin{tcolorbox}
(1) We have gathered papers utilizing over 30 distinct LLMs for cybersecurity tasks. These LLMs have been categorized into three groups based on their underlying architecture or principles: encoder-only, encoder-decoder, and decoder-only LLMs.

(2) We analyzed the trend in employing LLMs for cybersecurity tasks, revealing that decoder-only architectures are the most prevalent. Specifically, over 15 LLMs fall into the decoder-only category, and a total of 133 papers have investigated the utilization of decoder-only LLMs in cybersecurity tasks.

(3) Our analysis reveals a clear trend of \textbf{task-architecture alignment}. While decoder-only models have become dominant since 2023 for their generative and reasoning capabilities, encoder-only and encoder-decoder architectures remain highly relevant for specific niches. Encoder-only models are consistently chosen for representation-heavy classification tasks, while encoder-decoder models are often preferred for structured sequence-to-sequence tasks like program repair.
\end{tcolorbox}

\section{RQ3: What domain specification techniques are used to adapt LLMs to security tasks?}~\label{rq3}

LLMs have demonstrated their efficacy across various intelligent tasks~\cite{kaddour2023challenges}. Initially, these models undergo pre-training on extensive unlabeled corpora, followed by fine-tuning for downstream tasks. However, discrepancies in input formats between pre-training and downstream tasks pose challenges in leveraging the knowledge encoded within LLMs efficiently. The techniques employed with LLMs for security tasks can be broadly classified into three categories: prompt engineering, fine-tuning, and external augmentation. We will delve into a comprehensive analysis of these three categories and further explore their subtypes, as well as summarize the connections between LLM techniques and various security tasks.

\subsection{Fine-tuning LLMs for Security Tasks}\label{finetuning}
Fine-tuning techniques are extensively utilized across various downstream tasks in NLP~\cite{sun2020finetuning4}, representing a crucial step in adapting pre-trained LLMs, whose initial training might not sufficiently cover the nuances of specialized domains like cybersecurity. This process involves further training the model on task-relevant datasets to adjust its parameters, aligning its capabilities more closely with specific security objectives~\cite{dodge2020finetuning1,qi2023finetuning2}. The extent and nature of fine-tuning often depend upon task complexity and dataset size~\cite{dodge2020finetuning1,qi2023finetuning2}. Importantly, fine-tuning can mitigate the constraints posed by model size, enabling smaller models fine-tuned for specific tasks to potentially outperform larger, general-purpose models lacking task-specific adaptation~\cite{khare2023vulde11, zheng2024finetuning3}. In our surveyed literature (185 papers), fine-tuning was employed in 57 studies (~31\%). Its application varies significantly across domains: it was most prevalent in information and content security (10 out of 20 papers, 50\%) and software and system security (43 out of 119 papers, ~36\%), with moderate use in network security (4 out of 26 papers, ~15\%). Notably, no papers applying fine-tuning were identified in the hardware security or blockchain security domains within our selection.

The primary motivation for fine-tuning in LLM4Security is to imbue general-purpose LLMs with domain-specific knowledge and capabilities, tailoring them for specific security tasks. For example, in software vulnerability management, models are fine-tuned to perform classification or generation. RealVul fine-tunes models like CodeBERT and CodeT5 on augmented datasets of PHP code snippets to improve vulnerability classification~\cite{web_vulde1}. VulLLM employs multi-task instruction fine-tuning on models including CodeBERT and CodeT5, training them simultaneously for vulnerability detection, localization, and explanation to enhance generalization~\cite{du-etal-2024-vulde22}. For vulnerability repair, fine-tuning on datasets containing vulnerable code and patches allows models like T5~\cite{bugde9} or other code LLMs~\cite{huang2023bugrepair10, wang2023bugrepair16, vulrepair1} to learn repair patterns. VulAdvisor further adapts this by fine-tuning CodeT5 with context-aware attention to generate natural language repair suggestions instead of direct code patches~\cite{vulrepair13}. Similarly, in information security, fine-tuning is applied to tasks like harmful content detection; EXPLAINHM fine-tunes a Flan-T5 model to classify harmful memes based on debates generated by a larger multimodal LLM~\cite{harmful8}. Fine-tuning is also used for adapting models to specific data types like network traffic~\cite{ids2} or system logs~\cite{karlsen2023log3} for anomaly detection. This adaptation process typically involves preparing a high-quality dataset representative of the target task and using established training frameworks~\cite{zheng2024finetuning3} to update the model's parameters.

Different strategies exist regarding the scope of parameter updates during fine-tuning. Full fine-tuning adjusts all parameters of the LLM, offering the potential for high adaptability and performance, particularly when the target task differs significantly from the pre-training objectives~\cite{lv2023fullfine1}. This approach is often employed when feasible, for instance, in tasks like bug repair~\cite{huang2023bugrepair10, wang2023bugrepair16, 10.1145/bugrepair4, paul2023bugrepair20} or vulnerability repair~\cite{vulrepair1, vulrepair4, vulrepair5} using dedicated datasets. However, full fine-tuning demands significant computational resources. Alternatively, Parameter-Efficient Fine-Tuning (PEFT) methods update only a small subset of parameters or add trainable adapters, significantly reducing computational and memory costs~\cite{shen2021partfine2}. PEFT techniques observed in LLM4Security include adapter-tuning~\cite{houlsby2019partfine3, 10.1145/bugde5, yang2023bugde8}, prompt-tuning~\cite{lester2021partfine4, chen2023vulde4}, and Low-Rank Adaptation (LoRA)~\cite{hu2021lora, silva2024bugrepair18}, alongside API-based fine-tuning for closed-source models~\cite{partfine1, hu2023harmful2, khare2023vulde11, deng2023progfuzz1, jin2023bugde2, ali2023ids9}. These PEFT methods make fine-tuning more accessible, especially for very large models or in resource-constrained environments, while still achieving effective task adaptation.

Beyond standard supervised fine-tuning, Reinforcement Learning (RL) based strategies are gaining attention for aligning LLMs more closely with specific security objectives or leveraging their outputs to guide automated processes~\cite{ouyang2022insgpt,deepseekr1,reinforcement}. These approaches often involve optimizing an agent's policy using reward signals derived from task execution or human feedback, sometimes incorporating LLM-generated insights into the reward function~\cite{bugrepair33}. For instance, in web application security, RL can fine-tune an LLM to generate more effective web fuzzing inputs by rewarding payload sequences that successfully bypass defenses or trigger desired responses~\cite{webfuzz2}. Similarly, RL techniques can enhance log analysis by optimizing models based on the accuracy or utility of the identified anomalies~\cite{han2023log5}. In the domain of fuzzing complex systems like JavaScript engines, CovRL utilizes coverage-guided RL: it processes execution coverage using methods like TF-IDF to assign higher importance to rarely executed paths, then uses this weighted coverage to generate a reward signal, thereby fine-tuning an LLM mutator to produce mutations more likely to explore these less-covered, potentially bug-rich areas~\cite{progfuzz13}. For automated program repair, RePair leverages RL with process-based feedback; instead of just learning from fixed code examples, the LLM's policy for generating patches is directly rewarded based on the outcome of executing the generated code against test suites, thus encouraging the model to learn repair strategies that lead to functional correctness~\cite{zhao-etal-2024-bugrepair30}. In a different application connecting LLMs and RL, CrashTranslator utilizes LLMs first to analyze stack traces and compute semantic similarities with UI elements; this information then shapes the reward function for an RL agent that learns an optimized UI exploration policy to effectively reproduce mobile application crashes~\cite{bugrepair26}.

Despite its benefits, fine-tuning LLMs for security tasks faces persistent challenges. Creating high-quality, large-scale, and representative labeled datasets remains a primary bottleneck; cybersecurity data is often scarce, imbalanced, noisy, or may contain inherent biases reflecting only known threats or specific environments, which can limit the model's real-world applicability~\cite{dodge2020finetuning1}. Furthermore, while generally less computationally intensive than pre-training, fine-tuning still demands considerable resources, particularly for full parameter updates on large models~\cite{han2024finetuning8}. Performance limitations also exist. There is a significant risk of overfitting to the specific patterns present in the fine-tuning dataset, which can impair the model's ability to generalize to novel or unseen vulnerability types, attack vectors, or code constructs encountered in practice~\cite{montesinos2022finetuning6}. Relatedly, catastrophic forgetting can occur, where the model loses some of its general language or reasoning capabilities acquired during pre-training~\cite{luo2023finetuning7}. A critical concern, especially relevant to security, is that fine-tuning can inadvertently compromise safety alignments embedded during pre-training, potentially making models more susceptible to generating insecure code or harmful content if not carefully managed~\cite{du2024finetuning5, qi2023finetuning2}.

\subsection{Prompting LLMs for Security Tasks}
Recent studies in natural language processing highlight the significance of prompt engineering~\cite{liu2021prompt1} as an emerging fine-tuning approach aimed at bridging the gap between the output expectations of LLMs during pretraining and downstream tasks. This strategy has demonstrated notable success across various NLP applications. Incorporating meticulously crafted prompts as features in prompt engineering has emerged as a fundamental technique for enriching interactions with LLMs like ChatGPT, Bard, among others. These customized prompts serve a dual purpose: they direct the LLMs towards generating specific outputs while also serving as an interface for tapping into the vast knowledge encapsulated within these models.

In prompt engineering, utilizing inserted prompts to provide task-specific knowledge is especially beneficial for security tasks with limited data features. This becomes crucial when conventional datasets (such as network threat reports, harmful content on social media, code vulnerability datasets, etc.) are restricted or do not offer the level of detail needed for particular security tasks. For example, in handling cyber threat analysis tasks~\cite{siracusano2023cti4}, one can construct prompts by incorporating the current state of the network security posture. This prompts LLMs to learn directly from the flow features in a zero-shot learning manner~\cite{xian2020zeroshot}, extracting structured network threat intelligence from unstructured data, providing standardized threat descriptions, and formalized categorization. In the context of program fuzzing tasks~\cite{hu2023progfuzz3}, multiple individual test cases can be integrated into a prompt, assisting LLMs in learning the features of test cases and generating new ones through few-shot learning~\cite{bendre2020fewshot}, even with limited input. For tasks such as penetration testing~\cite{deng2023pentest1} and hardware vulnerability verification~\cite{saha2023hardrepair3}, which involve multiple steps and strict logical reasoning relationships between steps, one can utilize a chain of thought (COT)~\cite{wei2023cot} to guide the customization of prompts. This assists LLMs in process reasoning and guides them to autonomously complete tasks step by step.

In LLM4Security, almost all security tasks listed in Table\ref{tab2} involve prompt engineering, highlighting the indispensable role of prompts. In conclusion, recent research emphasizes the crucial role of prompt engineering in enhancing the performance of LLMs for targeted security tasks, thereby aiding in the development of automated security task solutions.

\subsection{External Augmentation}

\begin{table}[]
\caption{External augmentation techniques involved in prior studies.}
\label{tab4}
\resizebox{0.99\linewidth}{!}{
\begin{tabular}{c|p{3.5cm}|p{3.5cm}|p{1.5cm}}
\hline
\textbf{Augmentation technique} & \multicolumn{1}{c|}{\textbf{Description}}                                                                                          & \multicolumn{1}{c|}{\textbf{Examples}}                                                                           & \multicolumn{1}{c}{\textbf{Reference}} \\ \hline
\multirow{4}{*}{Features augmentation}           & Incorporating task-relevant features implicitly present in the dataset into prompts.                                               & Adding bug descriptions, bug locations, code context or resampling for imbalanced traffic.                       &
\cite{jin2023bugde2}~\cite{zhang2023progfuzz2} \linebreak \cite{xia2024progfuzz4}~\cite{jamal2023phishing4}\linebreak
\cite{ali2023ids9}~\cite{zhang2023bugrepair17} \linebreak
\cite{liu2023vulde13}\\ \hline
\multirow{6}{*}{External retrieval}              & Retrieving task-relevant information available in external knowledge bases as input.                                               & An external structured corpus of network threat intelligence,a hybrid patch retriever for fix pattern mining.    & \cite{du2023bugde6}~\cite{wang2023bugrepair16}\linebreak
\cite{perrina2023cti5}~\cite{pro_fuzzing2}\linebreak
\cite{vulde19}~\cite{bugrepair24}\linebreak
\cite{vulrepair12}~\cite{shan2024log6}\linebreak
\cite{bugrepair23}~\cite{contract5}  \\ \hline
\multirow{7}{*}{External tools}                  & Analysis results from specialized tools serving as auxiliary inputs or validate the result with external tools.                                                               & Static analysis tools, penetration testing tools, Rule-Based Reasoning Engine                                                           &  \cite{happe2023pentest5}~\cite{alrashedy2024vulrepair6} \linebreak
\cite{armengolestapé2024reverse4}~\cite{web_vulde1}\linebreak
\cite{seccode2}~\cite{reverse8}\linebreak
\cite{vulrepair15}~\cite{bugrepair24}\linebreak
\cite{vulrepair13}~\cite{bugrepair26}\linebreak
\cite{contract8}~\cite{ahmad2023hardrepair1} \\ \hline
\multirow{4}{*}{Task-adaptive training}          & Different training strategies from pre-training to enhance the model's adaptability to the task.                                   & Contrastive learning, transfer learning, distillation.                                   &   
    
\cite{bugde4}~\cite{10.1145/bugde5}\linebreak \cite{vulrepair5}~\cite{pei2024reverse5}\linebreak \cite{han2023log5}~\cite{webfuzz2} \linebreak 
\cite{tang2023bugrepair14} ~\cite{cai2024harmful3}\\ \hline
\multirow{5}{*}{Inter-model interaction}         & \multirow{5}{*}Introducing multiple models (which can be LLMs or other models) to collaborate and interact.                                       & Multiple LLMs collaboration, graph neural networks, gGAN                                                      &    \cite{cai2024harmful3}~\cite{yang2023progfuzz6}\linebreak \cite{tang2023bugrepair14}~\cite{seccode2}\linebreak
\cite{bugrepair26}~\cite{bugrepair32}\linebreak
\cite{contract10}                               \\ \hline
\multirow{4}{*}{Rebroadcasting}                  & Applicable to multi-step tasks, broad-casting the output results of each step iteratively as part of the prompt for the next step. & Difficulty-based patch example replay, variables' name propagation                                               &   \cite{xu2023reverse2}~\cite{10.1145/bugrepair4}\linebreak
\cite{reverse8}                                     \\ \hline
\multirow{4}{*}{Post-process}                    & Customizing special processing strategies for LLMs' outputs to better match task requirements.                                     & Post-processing based on Levenshtein distance to mitigate hallucinations, formal verification for generated code &  \cite{chen2023vulde4}~\cite{tihanyi2023vulde14}\linebreak
\cite{ads2}~\cite{bugrepair27}\linebreak
\cite{contract9}~\cite{contract10}
\\ \hline
\multirow{5}{*}{Workflow Integration}                    & Integrating LLMs as components within specific task workflows, incorporating various forms of interactive feedback.                                    & LLM-guided fuzzing strategy generation, Self-check reasoning, Coverage-guided feedback loop, Integrate LLM outputs into the human rating &  \cite{pro_fuzzing1}~\cite{pro_fuzzing2}\linebreak
\cite{progfuzz9}~\cite{progfuzz10}\linebreak
\cite{harmful7}~\cite{harmful8}

\\ \hline
\end{tabular}
}
\end{table}

While LLMs undergo thorough pre-training on extensive datasets, employing them directly for tackling complex tasks in security domains faces numerous challenges due to the diversity of domain data, the complexity of domain expertise, and the specificity of domain goals~\cite{ling2023external1}. Several studies in LLM4Security introduce external augmentation methods to enhance the application of LLMs in addressing security issues. These external augmentation techniques facilitate improved interaction with LLMs, bridging gaps in their knowledge base, and maximizing their capability to produce dependable outputs based on their existing knowledge.

We summarized the external augmentation techniques combined with LLMs in previous studies, as shown in Table~\ref{tab4}, identifying 8 distinct categories. The first augmentation technique is feature augmentation. The effectiveness of LLMs in handling downstream tasks heavily relies on the features included in the prompts. We have observed that many studies employing LLMs for security tasks extract contextual relationships or other implicit features from raw data and integrate them with the original data to customize prompts. These implicit features encompass descriptions of vulnerabilities~\cite{zhang2023bugrepair17}, bug locations~\cite{jin2023bugde2}, threat flow graphs~\cite{liu2023vulde13}, and more. Incorporating these implicit features alongside raw data leads to enhanced performance compared to constructing prompts solely from raw data. The next augmentation technique is external retrieval. External knowledge repositories can mitigate the hallucinations or errors arising from the lack of domain expertise in LLMs. LLMs can continually interact with external knowledge repositories during pipeline processing and retrieve knowledge relevant to security tasks to provide superior solutions~\cite{perrina2023cti5}. Next are external tools; rule-based external tools can serve as specialized external knowledge repositories, and in addressing security tasks, LLMs can utilize results from external tools (like static analyzers or formal methods) to rectify their outputs, thereby avoiding redundancy and errors\cite{happe2023pentest5,alrashedy2024vulrepair6,xu2025directedgreyboxfuzzinglarge}. The fourth augmentation technique is task-adaptive training. Existing studies adopt various training strategies (beyond standard fine-tuning discussed in~\autoref{finetuning}), like contrastive learning or distillation, to strengthen LLMs' adaptability to complex security tasks, enabling them to generate more targeted outputs~\cite{chen2020contrast,tang2023bugrepair14,bugde4}. The fifth augmentation technique, inter-model interaction, has garnered significant attention when a single LLM may struggle to handle complex and intricate tasks. Decomposing the pipeline process and introducing multiple LLMs for enhanced performance have been explored~\cite{yang2023progfuzz6}. This approach leverages collaboration and interaction among models to harness the underlying knowledge base advantages of each LLM. When a single interaction is insufficient to support LLMs in tasks such as variable name recovery or generating complex program patches ~\cite{xu2023reverse2,10.1145/bugrepair4}, rebroadcasting is used: constructing prompts for LLMs multiple times continuously where the output results of each step are broadcast iteratively as part of the prompt for the next step. This helps reinforce the contextual relationship between each interaction, thereby reducing error rates. The next augmentation technique is post-processing, where LLMs' outputs are validated or processed for certain security tasks requiring specific types of output~\cite{tihanyi2023vulde14}. This process helps mitigate issues such as hallucinations arising from the lack of domain knowledge in LLMs~\cite{chen2023vulde4}. The last strategy is workflow integration, which involves integrating LLMs as components within specific task workflows, incorporating various forms of interactive feedback. This feedback, such as coverage metrics guiding fuzzing strategies~\cite{progfuzz9, progfuzz10}, self-check reasoning steps~\cite{progfuzz9}, or outputs being integrated into human rating systems~\cite{harmful7}, allows the workflow's context and objectives to steer the LLM's behavior.

External augmentation techniques have significantly boosted the effectiveness of LLMs across various security tasks, yielding competitive performance. From these studies, it is evident that external augmentation techniques have the potential to address issues such as hallucinations and high false positive rates caused by deficiencies in LLMs' domain knowledge and task alignment. We believe that the integration of LLMs with external techniques will be a trend in the development of automated security task solutions.

\tcbset{
  title=RQ3 - Summary,  
}

\begin{tcolorbox}
(1) We summarize the domain-specific techniques used in previous research to apply LLMs to security tasks, including prompt engineering, fine-tuning, and external augmentation.

(2) Prompt engineering is the most widely used domain technique, with almost all 185 papers employing this approach. Fine-tuning techniques were used in 31\% of the papers, while task-specific external augmentation techniques were adopted in 36\% of the papers.

(3) We categorize and discuss the fine-tuning and external augmentation techniques mentioned in these papers, and analyze their relevance to specific security tasks.

(4) A key insight is that state-of-the-art LLM4Security applications are increasingly designed as hybrid systems. While prompt engineering is universal, the most sophisticated approaches specialize the model's capabilities, either through internal adaptation (fine-tuning) or by integrating it with external data and tools (external augmentation). This indicates a clear trend towards using LLMs as a core reasoning engine within a larger, more reliable security workflow, rather than as standalone oracles.
\end{tcolorbox}
\section{RQ4: What is the difference in data collection and pre-processing when applying LLMs to security tasks?}~\label{rq4}

Data plays a vital role throughout the model training process~\cite{zha2023data1}. Initially, collecting diverse and rich data is crucial to enable the model to handle a wide range of scenarios and contexts effectively. Following this, categorizing the data helps specify the model's training objectives and avoid ambiguity and misinterpretation. Additionally, preprocessing the data is essential to clean and refine it, thereby enhancing its quality. In this chapter, we examine the methods of data collection, categorization, and preprocessing as described in the literature.

\subsection{Data Collection}
Data plays an indispensable and pivotal role in the training of LLMs, influencing the model's capacity for generalization, effectiveness, and performance~\cite{sun2022data2}. Sufficient, high-quality, and diverse data are imperative to facilitate the model's comprehensive understanding of task characteristics and patterns, optimize parameters, and ensure the reliability of validation and testing. Initially, we explore the techniques employed for dataset acquisition. Through an examination of data collection methods, we classify data sources into four categories: open-source datasets, collected datasets, constructed datasets, and industrial datasets.

\begin{figure}[hbt]
\centering
\includegraphics[width=0.99\textwidth]{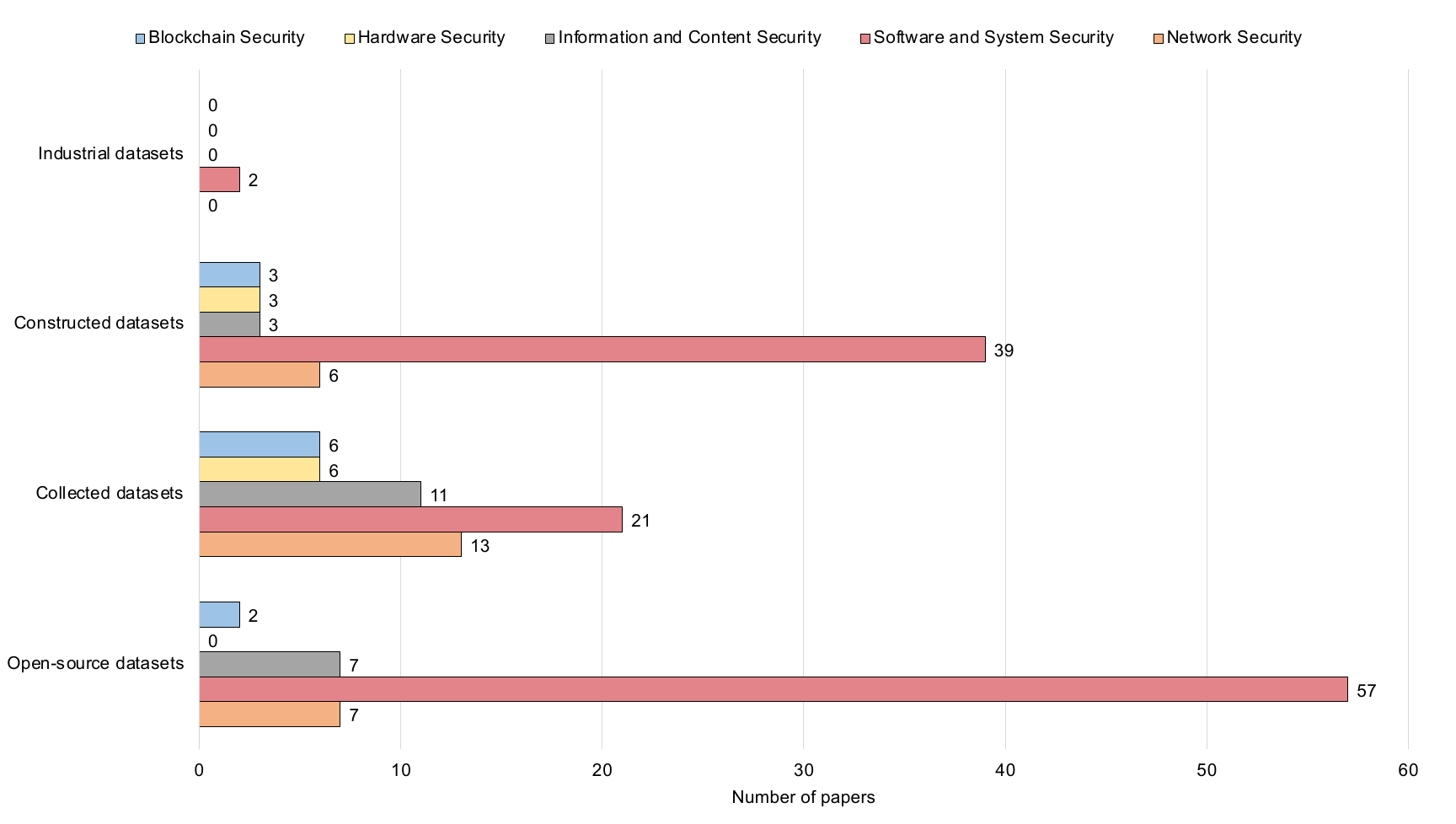}
\caption{The collection strategies of datasets in LLM4Security.}
\label{fig3}
\end{figure}

\textbf{Open-source datasets.} Open-source datasets refer to datasets that are publicly accessible and distributed through open-source platforms or online repositories~\cite{ids1,zhang2023vulde9,cai2024harmful3,chen2023contrast3}. For example, the UNSW-NB15 dataset contains 175,341 network connection records, including summary information, network connection features, and traffic statistics. The network connections in the dataset are labeled as normal traffic or one of nine different types of attacks~\cite{data3}. These are commonly used for benchmarking models or as a basis for fine-tuning on established tasks. The credibility of these datasets is ensured by their open-source nature, which also allows for community-driven updates. This makes them dependable resources for academic research.

\textbf{Collected datasets.} Researchers gather collected datasets directly from various sources, such as major websites, forums, blogs, and social media platforms. These datasets may include comments from GitHub, harmful content from social media, or vulnerability information from CVE websites, tailored to specific research questions. Such data is often curated specifically for training or evaluating models on niche or newly defined security problems.

\textbf{Constructed datasets.} The constructed dataset refers to a specialized dataset created by researchers  through the modification or augmentation of existing datasets to better suit their specific research goals~\cite{moskal2023cti6,vulrepair1,ahmad2023hardrepair1,koide2024phishing2}. These changes could be made through manual or semi-automated processes, which might entail creating test sets tailored to specific domains, annotating datasets, or generating synthetic data. Generated synthetic data, for instance, often serves to augment training sets for fine-tuning LLMs when real-world data is scarce (as discussed further in Section~\ref{rq3}). For instance, researchers might gather information on web vulnerabilities and the corresponding penetration testing methods, structure them into predefined templates to form vulnerability scenarios, which can then be used for evaluating LLM reasoning capabilities via prompting~\cite{deng2023pentest1}.

\textbf{Industrial datasets.} Industrial datasets are data obtained from real-world commercial or industrial settings, typically consisting of industrial applications, user behavior logs, and other sensitive information~\cite{zhang2023progfuzz2,liu2023log1}. These datasets are particularly valuable for research aimed at training and evaluating models intended for real-world deployment and application scenarios.

The Figure~\ref{fig3} illustrates the data collection strategies for LLM-related datasets. From the data depicted in the Figure~\ref{fig3}, it can be observed that 73 studies utilize open-source datasets. The utilization of open-source datasets is predominantly attributed to their authenticity and credibility. These datasets are typically comprised of real-world data sourced from diverse origins, including previous related research, thereby ensuring a high degree of reliability and stability to real-world scenarios. This authenticity enables LLMs to learn from genuine examples, facilitating a deeper understanding of real-world security tasks and ultimately improving their performance when used for fine-tuning or evaluation. Additionally, due to the recent emergence of LLMs, there is indeed a challenge of the lack of suitable training sets. Hence, researchers often collect data from websites or social media (collected datasets) and construct datasets to serve as specific training, fine-tuning, or evaluation sets to make the data more suitable for specific security tasks. We also analyzed the relationship between data collection strategies and the security domain. In certain domains such as network security, the preference for collecting datasets surpasses that of using open-source datasets. This indicates that obtaining data for applying LLMs to certain security tasks is still inconvenient. Among the 185 papers examined, only 2 studies utilized industrial datasets. This indicates a potential gap between the characteristics of datasets used in academic research and those in real-world industrial settings. This difference underscores the importance of future research exploring industrial datasets to ensure the applicability and robustness of LLMs across academic and industrial domains. Some papers focus on exploring the use of existing LLMs, such as ChatGPT, in security tasks~\cite{10.1007/hardrepair2,cryptoeprint:2023/212hardrepair5}. These papers often do not specify the datasets used for model training, as LLMs like ChatGPT typically do not require users to prepare their own training data for application scenarios but are instead guided via prompting or in-context learning.

\subsection{Types of Datasets}

The choice of data types plays a crucial role in shaping the architecture and selection of LLMs, as they directly influence the extraction of implicit features and subsequent decision-making by the model during both training and inference. This decision significantly impacts the overall performance and ability of LLMs to generalize~\cite{liu2024data4}. We conducted a thorough analysis and categorization of the data types utilized in LLM4Security research. Through examining the interplay between data types, model architectures, and task demands, our goal is to highlight the vital significance of data types in effectively applying LLMs to security-related tasks.

\begin{table}[]
\caption{Data types of datasets involved in prior studies.}
\label{tab5}
\resizebox{0.99\linewidth}{!}{
\begin{tabular}{r|l|c|c|p{3cm}}
\hline
\textbf{Category}                     & \textbf{Data type}                             & \textbf{Studies} & \textbf{Total}       & \textbf{References} \\ \hline
\multirow{23}{*}{Code-based datasets} & Vulnerable code                                & 24               & \multirow{23}{*}{90} & \cite{chen2023vulde1}~\cite{ferrag2023vulde2}~\cite{chen2023vulde4}~\cite{chow2023vulde6}~\linebreak~\cite{liu-etal-2023vulde7}~\cite{vulde8}~\cite{zhang2023vulde9}~\cite{zhang2023vulde10}~\linebreak~\cite{khare2023vulde11}~\cite{ullah2023vulde12}~\cite{liu2023vulde13}~\cite{vulrepair1}~\linebreak~\cite{charalambous2023vulrepair3}~\cite{alrashedy2024vulrepair6}~\cite{ahmad2023bugde1}~\cite{chen2023contrast3}~\cite{ullah2024vulde17}~\cite{web_vulde1}~\cite{vulde21}~\cite{vulrepair15}~\cite{contract7}\linebreak\cite{contract9}~\cite{contract10}~\cite{contract8} \\ \cline{2-3} \cline{5-5} 
                                      & Source code                                    & 23               &                      &\cite{bakhshandeh2023vulde3}~\cite{hu2023progfuzz3}~\cite{progfuzz5}~\cite{yang2023progfuzz6}~\linebreak~\cite{xu2023reverse2}~\cite{reverse3}~\cite{armengolestapé2024reverse4}~\cite{pei2024reverse5}~\linebreak~\cite{reverse6}~\cite{liu2023vulde13}~\cite{sun2023contrast1}~\cite{david2023contrast2}~\linebreak~\cite{hu2023contrast4}~\cite{gai2023blockchainsec1}~\cite{Hu2024reverse7}~\cite{obfuscation1}~\cite{progfuzz10}~\cite{progfuzz13}~\cite{malware4}\linebreak\cite{reverse9}~\cite{bugde10}~\cite{reverse10}~\cite{hardwareip1}                  \\ \cline{2-3} \cline{5-5} 
                                      & Bug-fix pairs                                  & 16               &                      &\cite{jin2023bugde2}~\cite{li2023bugde7}~\cite{zhang2023bugrepair2}~\cite{10.1145/bugrepair4}~\linebreak~\cite{paul2023bugrepair20}~\cite{olausson2024bugrepair7}~\cite{bugrepair9}~\cite{huang2023bugrepair10}~\linebreak~\cite{zhang2023bugrepair11}~\cite{xia2023bugrepair12}~\cite{xia2023bugrepair13}~\cite{wang2023bugrepair16}~\linebreak~\cite{zhang2023bugrepair17}~\cite{silva2024bugrepair18}~\cite{bugrepair23}~\cite{bugrepair25}                     \\ \cline{2-3} \cline{5-5} 
                                      & Buggy code                                           & 10               &                      &\cite{li2023bugde3}~\cite{10.1145/bugde5}~\cite{sobania2023bugrepair21}~\cite{paul2023bugrepair20}~\linebreak~\cite{huang2023bugrepair10}~\cite{deng2023progfuzz1}~\cite{ahmad2023hardrepair1}~\cite{bugrepair22}~\cite{zhao-etal-2024-bugrepair30}~\cite{wang-etal-2024-bugde11}                    \\ \cline{2-3} \cline{5-5} 
                                      & Traffic packages                               & 4                &                      &\cite{ids4}~\cite{ids2}~\cite{ids1}~\cite{ids6}                    \\ \cline{2-3} \cline{5-5} 
                                      & Patches                                        & 3                &                      &\cite{khare2023vulde11}~\cite{bugrepair5}~\cite{tang2023bugrepair14}                     \\ \cline{2-3} \cline{5-5} 
                                      & Code changes                                   & 3                &                      &\cite{yang2023bugde8}~\cite{du2023bugde6}~\cite{bugrepair8}                    \\ \cline{2-3} \cline{5-5} 
                                      & Vulnerability-fix pairs                        & 3                &                      &\cite{vulrepair5}~\cite{vulrepair4}~\cite{vulrepair13}                   \\ \cline{2-3} \cline{5-5} 
                                      & Web attack payloads                            & 2                &                      &\cite{webfuzz1}~\cite{webfuzz2}                    \\ \cline{2-3} \cline{5-5} 
                                      &Subject protocol programs                            & 1                &                      &\cite{meng2024webfuzz4}                   \\ \cline{2-3} \cline{5-5} 
                                      & Vulnerable programs                            & 1                &                      &\cite{vulrepair2}                     \\ \hline
\multirow{26}{*}{Text-based datasets} & Prompts                                        & 19               & \multirow{26}{*}{60}&\cite{ali2023ids9}~\cite{moskal2023cti6}~\cite{deng2023pentest1}~\cite{temara2023pentest3}~\linebreak~\cite{charan2023pentest4}~\cite{happe2023pentest5}~\cite{bugrepair3}~\cite{tol2023vulrepair10}~\linebreak~\cite{zhang2023progfuzz2}~\cite{pearce2022reverse1}~\cite{malware2}~\cite{heiding2023phishing1}~\linebreak~\cite{roy2024phishing3}~\cite{harmful6}~\cite{Forensics1}~\cite{saha2023hardrepair3}~\linebreak~\cite{lin2023hardrepair4}~\cite{seccode2}~\cite{contract5}  \\ \cline{2-3} \cline{5-5} 
                                      & Log messages                                   & 8                &                      & \cite{liu2023log1}~\cite{chen2023log2}~\cite{karlsen2023log3}~\cite{qi2023log4}~\linebreak~\cite{han2023log5}~\cite{shan2024log6}~\cite{log7}~\cite{log8}                  \\ \cline{2-3} \cline{5-5} 
                                      & Social media contents                           & 7                &                      & \cite{hanley2023harmful1}~\cite{hu2023harmful2}~\cite{cai2024harmful3}~\cite{mets2023harmful4}~\linebreak~\cite{steganography1}~\cite{harmful7}~\cite{harmful8}                    \\ \cline{2-3} \cline{5-5} 
                                      & Spam messages                                  & 4                &                      & \cite{labonne2023phishing5}~\cite{phishing7}~\cite{cambiaso2023phishing8}~\cite{jamal2023phishing4}                    \\ \cline{2-3} \cline{5-5} 
                                      & Bug reports                                    & 4                &                      &\cite{bugde4}~\cite{10.1145/bugde5}~\cite{du2023bugde6}~\cite{bugrepair27}                     \\ \cline{2-3} \cline{5-5} 
                                      & Accident reports                            & 2                &                      &\cite{ads1}~\cite{ads2}                   \\ \cline{2-3} \cline{5-5} 
                                      & Attack descriptions                            & 2                &                      &\cite{ids7}~\cite{ids3}                    \\ \cline{2-3} \cline{5-5} 
                                      & CVE reports                                    & 2                &                      &\cite{aghaei2023ids5}~\cite{aghaei2023cti3}                     \\ \cline{2-3} \cline{5-5}
                                       & Commit messages                             & 2                &                      &\cite{zhang2023bugrepair2}~\cite{wang2023bugrepair16}                     \\ \cline{2-3} \cline{5-5} 
                                      & Cyber threat intelligence data                 & 2                &                      &\cite{phishing6}~\cite{koide2024phishing2}                     \\ \cline{2-3} \cline{5-5} 
                                      & Protocol Messages
                                        & 2                &                      &\cite{pro_fuzzing1}~\cite{pro_fuzzing2}                     \\ \cline{2-3} \cline{5-5} 
                                        & Program documentations                         & 2                &                      &\cite{xia2024progfuzz4}~\cite{progfuzz9}                     \\ \cline{2-3} \cline{5-5} 
                                      & Top-level domains                              & 1                &                      &\cite{liu2023ids8}                     \\ \cline{2-3} \cline{5-5} 
                                      & Cellular network specifications                & 1                &                      &\cite{pro_analysis1}                     \\ \cline{2-3} \cline{5-5} 
                                      & Security reports                               & 1                &                      & \cite{cti1}                    \\ \cline{2-3} \cline{5-5} 
                                      & Threat reports                                 & 1                &                      &\cite{siracusano2023cti4}                     \\ \cline{2-3} \cline{5-5} 
                                      & Structured threat information                  & 1                &                      &\cite{perrina2023cti5}                     \\ \cline{2-3} \cline{5-5} 
                                      & Antivirus scan reports                         & 1                &                      &\cite{joyce2023malware1}                     \\ \cline{2-3} \cline{5-5} 
                                      & Passwords                                      & 1                &                      &\cite{rando2023access2}                     \\ \cline{2-3} \cline{5-5} 
                                      & Hardware documentations                        & 1                &                      &\cite{meng2023hardde1}                     \\ \hline
\multirow{10}{*}{Combined datasets}                     & Vulnerable code and vulnerability descriptions & 7                & \multirow{10}{*}{23}                    & \cite{liu-etal-2023vulde7}~\cite{chen2023vulde4}~\cite{seccode2}~\cite{vulde20}~\cite{vulrepair12}~\cite{vulrepair14}~\cite{contract6}  
\\ \cline{2-3} \cline{5-5}
& Buggy code and bug reports & 4                &           & \cite{bugrepair24}~\cite{bugrepair26}~\cite{wang-etal-2024-bugrepair31}~\cite{bugrepair33} 
\\ \cline{2-3} \cline{5-5}
& Vulnerability-fix pairs and vulnerability descriptions & 3                &           & \cite{xu2024vulde18}~\cite{du-etal-2024-vulde22}~\cite{ahmad2023hardrepair1} 
\\ \cline{2-3} \cline{5-5}
& Bug-fix pairs and bug reports & 3                &           & \cite{bugrepair28}~\cite{bugrepair29}~\cite{bugrepair32} 
\\ \cline{2-3} \cline{5-5}
& Vulnerability-fix pairs and repair strategies & 1                &           & \cite{vulrepair11} 
\\ \cline{2-3} \cline{5-5}
& Code changes and vulnerability descriptions & 1                &           & \cite{vulrepair14} 
\\ \cline{2-3} \cline{5-5}
& Source code and program documentations & 1                &           & \cite{zhang2023progfuzz2} 
\\ \cline{2-3} \cline{5-5}
& Buggy code and commit messages & 1               &           & \cite{bugde9}
\\ \cline{2-3} \cline{5-5}
& Assertion code and hardware module descriptions & 1                &     & \cite{assertiongen1}
\\ \cline{2-3} \cline{5-5}
& Decompiled code and variable names & 1                &                    & \cite{reverse8}  
\\ \hline

\end{tabular}
}

\end{table}

\textbf{Data type categorization.}\  We categorize all datasets into three types: code-based, text-based, and combined data types. Table~\ref{tab5} provides a detailed breakdown of the specific data included in each category, derived from 185 studies. The analysis reveals that the majority of instances rely on code-based datasets, constituting a total of 90 instances. This dominance underscores the inherent code analysis capabilities of LLMs when applied for security tasks. These models demonstrate proficiency in understanding and processing code data, making code-based datasets suitable for fine-tuning models for tasks like vulnerability detection, program fuzzing, and bug repair. Their capacity to handle and learn from extensive code data enables LLMs to offer robust insights and solutions for various security applications either through fine-tuning or direct prompting.

Text-based datasets (total 60 instances) encompass a wide variety of textual information relevant to cybersecurity. While understanding the intricacies of training data might not be crucial for closed-source LLMs like ChatGPT, insights into data handling techniques for other models are still valuable. These datasets range from structured formats like prompts used for direct task execution~\cite{deng2023pentest1,liu2023vulde13} or logs used for fine-tuning anomaly detection models~\cite{liu2023log1,chen2023log2,karlsen2023log3}, to unstructured natural language found in bug reports~\cite{bugde4,du2023bugde6}, CVE reports~\cite{aghaei2023ids5,cti1}, commit messages~\cite{vulde19,bugrepair24}, security policies~\cite{seccode2}, hardware documentation~\cite{meng2023hardde1}, and social media content~\cite{hu2023harmful2,hanley2023harmful1}. Specific security tasks necessitate particular text data inputs, often utilized for fine-tuning specialized classifiers (e.g., for harmful content) or providing context for LLM analysis via prompting (e.g., CVE reports).

The prevalence of vulnerable code (34 instances), source code (23), and bug-fix pairs (16) in code-based datasets can be attributed to their ability to effectively meet task requirements for fine-tuning or evaluation. Vulnerable code naturally exhibits semantic features of code containing vulnerabilities to LLMs, thereby highlighting the distinguishing traits of vulnerable code when juxtaposed with normal code snippets when training detection models. A similar rationale applies to bug-fix pairs, which are essential for fine-tuning program repair models. Source code serves as the backbone of any software project, encompassing the logic and instructions that define program behavior. Thus, having a substantial amount of source code data is essential for training LLMs (either during pre-training or fine-tuning) to grasp the intricacies of programs, enabling them to proficiently generate, analyze, and comprehend code across various security tasks. Additionally, commonly used data types such as buggy code (10) and traffic packages (4), are also widespread for training detection or analysis models.

Combined datasets, integrating multiple data types, represent a growing approach (23 instances identified). This strategy often pairs code artifacts with relevant natural language text, such as combining vulnerable code with vulnerability descriptions~\cite{liu-etal-2023vulde7,chen2023vulde4}, buggy code with bug reports~\cite{bugrepair24,bugrepair26}, or source code with program documentation~\cite{zhang2023progfuzz2}. The key advantage lies in leveraging the LLM's ability to process both structured code and unstructured text simultaneously~\cite{wang2024planningnaturallanguageimproves}. By providing natural language context (e.g., descriptions of what a vulnerability entails or what a function should do), these combined datasets help LLMs achieve a deeper semantic understanding of the code during fine-tuning or prompting, leading to more accurate vulnerability detection, better code repair suggestions, or more effective code analysis compared to using code data alone.

\subsection{Data Pre-processing}

\begin{table}[]
\caption{The data preprocessing techniques for code-based datasets.}
\label{tab6}
\resizebox{0.99\linewidth}{!}{
\begin{tabular}{c|p{4.5cm}|p{4cm}|p{1.5cm}}
\hline
\multicolumn{1}{c|}{\textbf{Preprocessing techniques}} & \multicolumn{1}{c|}{\textbf{Description}}                                                                                                                                                                      & \multicolumn{1}{c|}{\textbf{Examples}}                                                                                                     & \textbf{References} \\ \hline
\multirow{4}{*}{Data extraction}                                        & Retrieve pertinent code segments from code-based datasets tailored to specific security tasks, accommodating various levels of granularity and specific task demands.                     & Token-level, statement-level, class-level, traffic flow.                                                                                                              &\cite{reverse3}~\cite{chan2023vulde16}~\linebreak~\cite{ids4}                  \\ \hline
\multirow{4}{*}{Duplicated instance deletion}                           & Eliminate duplicate instances from the dataset to maintain data integrity and avoid repetition during the training phase.                                           & Removal of duplicate code, annotations, and obvious vulnerability indicators in function names.                                            & \cite{vulde8}~\cite{zhang2023vulde9}~\linebreak~\cite{zhang2023bugrepair17} \\ \hline
\multirow{6}{*}{Unqualified data deletion}                              & Remove unfit data by implementing filtering criteria to preserve suitable samples, ensuring the dataset's quality and suitability for diverse security tasks. & Remove or anonymize conments and information that may provide obvious hints about the vulnerability~(package, variable names, and strings,etc.). &  \cite{ferrag2023vulde2}~\cite{khare2023vulde11}~\linebreak~\cite{10.1145/bugrepair4}~\cite{paul2023bugrepair20}                   \\ \hline
\multirow{2}{*}{Code representation}                                    & Represent code as tokens, code graph, AST, etc.                                                                                                                                                                                         & Tokenize source or binary code as tokens or transform code into call graph.                                                                                                  &\cite{zhang2023vulde10}~\cite{bugrepair9}~\linebreak~\cite{huang2023bugrepair10}~\cite{contract8}                    \\ \hline
\multirow{4}{*}{Data segmentation}                                      & Divide the dataset into training, validation, and testing subsets for model training, parameter tuning, and performance evaluation.                                                                                         & Partition the dataset based on specific criteria, which may include division into training, validation, or testing subsets.
&\cite{yang2023bugde8}~\cite{reverse6}                     \\ \hline
\end{tabular}
}
\end{table}

When training and using LLMs, it's important to preprocess the initial dataset to obtain clean and appropriate data for model training~\cite{10.1145/bugde5}. Data preprocessing involves tasks like cleaning, reducing noise, and normalization. Different types of data may require different preprocessing methods to improve the performance and effectiveness of LLMs in security tasks, maintaining data consistency and quality. This section will provide a detailed explanation of the data preprocessing steps customized for the two main types of datasets: those based on code and those based on text.

\textbf{Data preprocessing techniques for code-based datasets.} We outline the preprocessing techniques utilized for code-based datasets, often comprising five key steps. The initial step involves \textbf{data extraction}, retrieving relevant code snippets from diverse sources. Depending on the research task's needs~\cite{reverse3,ids4}, snippets may be extracted at different levels of detail, ranging from individual lines, methods, or functions to entire code files or projects. To prevent bias and redundancy during training, the next step removes \textbf{duplicated instances} by identifying and eliminating them from the dataset~\cite{zhang2023vulde9,zhang2023bugrepair17}, enhancing diversity and uniqueness. \textbf{Unqualified data deletion} follows, removing snippets that don't meet predefined quality standards (e.g., based on size, complexity, or presence of obvious hints) to ensure relevance to the security task and avoid noise~\cite{ferrag2023vulde2,10.1145/bugrepair4}. \textbf{Code representation} converts snippets into suitable formats for LLM processing, typically involving tokenization for security tasks~\cite{bugrepair9}. Finally, \textbf{data splitting} divides the preprocessed dataset into training, validation, and testing subsets~\cite{yang2023bugde8}. Training sets are used to train the LLM, validation sets help tune hyperparameters, and testing sets assess model performance on unseen data.

\textbf{Data preprocessing techniques for text-based datasets.} As depicted in Table~\ref{tab7}, preprocessing text-based datasets involves five steps, with minor differences compared to code-based datasets. The process begins with data extraction, carefully retrieving text from various sources such as bug reports~\cite{bugde4}, program documentation~\cite{xia2024progfuzz4}, hardware documentation~\cite{meng2023hardde1}, and social media content~\cite{hanley2023harmful1}. This initial phase ensures the dataset encompasses a range of task-specific textual information. After data extraction, the text undergoes segmentation tailored to the specific research task's needs. Segmentation may involve breaking text into sentences or further dividing it into individual words for analysis~\cite{aghaei2023cti3,meng2023hardde1}. Subsequent preprocessing operations standardize and clean the text, typically involving the removal of specific symbols, stop words, and special characters~\cite{mets2023harmful4,aghaei2023cti3}. This standardized textual format facilitates effective processing by LLMs. To address bias and redundancy in the dataset, this step enhances dataset diversity, aiding the model's generalization to new inputs~\cite{labonne2023phishing5}. Data tokenization is essential for constructing LLM inputs, where text is tokenized into smaller units like words or subwords to facilitate feature learning~\cite{aghaei2023cti3}. Finally, the preprocessed dataset is divided into subsets, typically comprising training, validation, and testing sets.

\begin{table}[]
\caption{The data preprocessing techniques for text-based datasets.}
\label{tab7}
\resizebox{0.99\linewidth}{!}{
\begin{tabular}{c|p{4.5cm}|p{4cm}|p{1.5cm}}
\hline
\textbf{Preprocessing techniques} & \multicolumn{1}{c|}{\textbf{Description}}                                                                                              & \multicolumn{1}{c|}{\textbf{Examples}}                                                                                  & \textbf{References}  \\ \hline
\multirow{3}{*}{Data extraction}                  & Retrieve appropriate text from documentation based on various security tasks. & Attack description, bug reports, social media content, hardware documentation, etc.                                         &\cite{ids3}~\cite{aghaei2023ids5}~\linebreak~\cite{xia2024progfuzz4}~\cite{hanley2023harmful1}~\linebreak~\cite{meng2023hardde1}                      \\ \hline
\multirow{2}{*}{Initial data segmentation}         & Categorize data into distinct groups as needed.                                          & Split data into sentences or words.                                                                                     &\cite{labonne2023phishing5}~\cite{mets2023harmful4}~\linebreak~\cite{aghaei2023cti3}  \\ \hline
\multirow{4}{*}{Unqualified data deletion}         & Delete invalid text data according to the specified rules.                                   & Remove certain symbols and words (rare words, stop words, etc.), or convert all content to lowercase.                     & \cite{bugde4}~\cite{cai2024harmful3}~\linebreak~\cite{cti1}                     \\ \hline
\multirow{2}{*}{Text representation}               & Token-based text representation.                                                                                                       & Tokenize the texts, sentences, or words into tokens.                                                                    &\cite{aghaei2023cti3}~\cite{meng2023hardde1}                     \\ \hline
\multirow{4}{*}{Data segmentation}                 &Divide the dataset into training, validation, and testing subsets for model training, parameter tuning, and performance evaluation.                                                                                  & Partition the dataset based on specific criteria, which may include division into training, validation, or testing subsets. &  \cite{rando2023access2}~\cite{steganography1}~\linebreak~\cite{liu2023ids8}                    \\ \hline

\end{tabular}

}
\end{table}

\subsection{Data Augmentation using LLMs}\label{sec:data_augmentation} 
A recurring challenge in applying LLMs to cybersecurity tasks is the scarcity of high-quality, labeled data. Datasets for vulnerabilities, malware, specific attack types, or other security events can be difficult to obtain in sufficient volume and diversity, hindering the training of robust models. To address this bottleneck, a number of studies have explored leveraging the generative capabilities of LLMs themselves for data augmentation~\cite{dodge2020finetuning1}. The overarching goal is to enhance dataset size, diversity, and balance, thereby improving the robustness and generalization performance of security models trained (typically via fine-tuning) on the augmented data.

These LLM-based data augmentation strategies manifest in various forms depending on the task and data type. One common approach involves generating synthetic code samples used to enlarge fine-tuning datasets. This includes creating new examples of vulnerable code, such as RealVul using code transformations based on PHP vulnerability patterns~\cite{web_vulde1}, prompting LLMs with CWE details to generate pairs of vulnerable and non-vulnerable functions~\cite{wang-etal-2024-bugde11}, implanting specific bug types into correct code snippets to create debugging benchmarks~\cite{bugrepair29}, or generating formally verified vulnerable and benign code snippets~\cite{tihanyi2023vulde14}. LLMs are also used to create specific malicious entities, for instance, generating diverse synthetic variations of malicious NPM packages based on known seeds to improve the fine-tuning of malware detection training~\cite{malware4}. In the textual domain, LLMs synthesize relevant data often used for fine-tuning or as few-shot examples, such as examples of harmful online discourse~\cite{harmful6}, natural language repair suggestions ("oracles") generated by prompting an LLM with vulnerable code and patches which then form a corpus for fine-tuning suggestion models~\cite{vulrepair13}, concise vulnerability explanations derived from patch features used for training auxiliary tasks during multi-task fine-tuning~\cite{du-etal-2024-vulde22}, or richer textual features like the explanatory 'debate text' produced by EXPLAINHM to fine-tune a classifier for harmful meme classification~\cite{harmful8}. In a related application, LLMs assist in augmenting or constructing the knowledge bases essential for security tools, exemplified by VULTURE, which uses LLMs to analyze CVEs and commits for building a comprehensive third-party library vulnerability database~\cite{xu2024vulde18} used by the downstream detection tool.

\tcbset{
  title=RQ4 - Summary,  
}

\begin{tcolorbox}
(1) Our analysis of data sources reveals a significant reliance on open-source, collected, and constructed datasets, reflecting a persistent challenge of data scarcity. A critical insight is the simultaneous disconnect from industrial practice, evidenced by the extreme scarcity of industrial datasets (used in only 1\% of papers), which raises questions about the real-world applicability of current models.

(2) We categorize datasets into three primary types: code-based, text-based, and combined. While code-based data is predominant, a key trend is the growing use of combined datasets, which pair code with natural language descriptions to provide crucial context and enhance the semantic understanding of LLMs.

(3) To address the data scarcity challenge, a notable insight is the trend of leveraging LLMs for data augmentation. Our review shows researchers are increasingly using LLMs to generate synthetic code samples, malicious entities, and diverse textual data to enhance the size and quality of training sets.

(4) We summarize the common data preprocessing pipeline applied across studies, which includes essential steps such as data extraction, cleaning and deduplication, converting data into suitable representations for LLMs, and segmentation into training, validation, and testing sets.
\end{tcolorbox}

\section{Threats to Validity}\label{sec:threats}

\textbf{Paper retrieval omissions.} One significant potential risk is the possibility of overlooking relevant papers during the search process. While collecting papers on LLM4Security tasks from various publishers, there is a risk of missing out on papers with incomplete abstracts, lacking cybersecurity tasks or LLM keywords. To address this issue, we employed a comprehensive approach that combines manual searching, automated searching, and snowballing techniques to minimize the chances of overlooking relevant papers as much as possible. We extensively searched for LLM papers related to security tasks in three top security conferences, extracting authoritative and comprehensive security task and LLM keywords for manual searching. Additionally, we conducted automated searches using carefully crafted keyword search strings on seven widely used publishing platforms. Furthermore, to further expand our search results, we employed both forward and backward snowballing techniques.

\textbf{Bias of research selection.} The selection of studies carries inherent limitations and potential biases. Initially, we established criteria for selecting papers through a combination of automated and manual steps, followed by manual validation based on Quality Assessment Criteria (QAC). However, incomplete or ambiguous information in BibTeX records may result in mislabeling of papers during the automated selection process. To address this issue, papers that cannot be conclusively excluded require manual validation. However, the manual validation stage may be subject to biases in researchers' subjective judgments, thereby affecting the accuracy of assessing paper quality. To mitigate these issues, we enlisted two experienced reviewers from the fields of cybersecurity and LLM to conduct a secondary review of the research selection results. This step aims to enhance the accuracy of paper selection and reduce the chances of omission or misclassification. By implementing these measures, we strive to ensure the accuracy and integrity of the selected papers, minimize the impact of selection biases, and enhance the reliability of the systematic literature review. Additionally, we provide a replication package for further examination by others.
\section{Challenges and Opportunities}\label{sec:challenges}
\subsection{Challenges}
\subsubsection{Challenges in LLM Applicability.} 
\
\newline\textbf{Model size and deployment.} The size of LLMs have seen significant growth over time, escalating from 117M parameters for GPT-1 to 1.5B parameters for GPT-2, and further to 175B parameters for GPT-3~\cite{yang2023challenge1}. Models with billions or even trillions of parameters present substantial challenges in terms of storage, memory, and computational demands~\cite{fedus2022challenge2}. This can potentially impede the deployment of LLMs, particularly in scenarios where developers lack access to potent GPUs or TPUs, especially in resource-constrained environments necessitating real-time deployment. CodeBERT~\cite{feng2020codebert} emerged in 2019 as a pre-trained model featuring 125M parameters and a model size of 476MB. Recent models like Codex~\cite{chen2021codex} and CodeGen~\cite{nijkamp2023codegen} have surpassed 100 billion parameters, with model sizes exceeding 100GB. Larger sizes entail more computational resources and higher time costs. For instance, training the GPT-Neox-20B model~\cite{Black2021GPTNeo} mandates 825GB of raw text data and deployment on 8 NVIDIA A100-SXM4-40GB graphics processing units (GPUs). Each GPU comes with a price tag of over \$6,000, and the training duration spans 1,830 hours or roughly 76 days. These instances underscore the substantial computational costs linked with training LLMs. Additionally, these platforms entail notable energy expenses, with LLM-based platforms projected to markedly amplify energy consumption~\cite{rillig2023challenge3}. Some vendors like OpenAI and Google provide online APIs for LLMs to alleviate user usage costs, while researchers explore methods to curtail LLM scale. Hsieh et al.~\cite{hsieh2023challenge4} proposed step-by-step distillation to diminish the data and model size necessary for LLM training, with their findings showcasing that a T5 model with only 770MB surpassed a 540B PaLM.

\textbf{Data Quality, Availability, and Security Challenges.}
In Section~\ref{rq4}, we conducted an extensive examination of the datasets and data preprocessing procedures employed in the 118 studies. Our analysis unveiled the heavy reliance of LLMs on a diverse array of datasets for training and fine-tuning. The findings underscore the challenge of data scarcity encountered by LLMs when tackling security tasks. The quality, diversity, and volume of data directly influence the performance and generalization capabilities of these models. Given their scale, LLMs typically necessitate substantial data volumes to capture nuanced distinctions, yet acquiring such data poses significant challenges. Many specific security tasks suffer from a dearth of high-quality and robust publicly available datasets. Relying on limited or biased datasets may result in models inheriting these biases, leading to skewed or inaccurate predictions. Furthermore, there is a concern regarding the risk of benchmark data contamination, where existing research may involve redundant filtering of native data, potentially resulting in overlap between training and testing datasets, thus inflating performance metrics~\cite{lee2022challenge5}. Additionally, we raise serious apprehensions regarding the inclusion of personally private information, such as phone numbers and email addresses, in training corpora when LLMs are employed for information and content security tasks, which precipitate privacy breaches during the prompting process~\cite{elmhamdi2023challenge6}.

\subsubsection{Challenges in LLM Generalization Ability.} The generalization capability of LLMs pertains to their ability to consistently and accurately execute tasks across diverse tasks, datasets, or domains beyond their training environment. Despite undergoing extensive pre-training on large datasets to acquire broad knowledge, the absence of specialized expertise can present challenges when LLMs encounter tasks beyond their pre-training scope, especially in the cybersecurity domain. As discussed in Section ~\ref{rq2}, we explored the utilization of LLMs in 21 security tasks spanning five security domains. We observed substantial variations in the context and semantics of code or documents across different domains and task specifications. To ensure LLMs demonstrate robust generalization, meticulous fine-tuning, validation, and continuous feedback loops on datasets from various security tasks are imperative. Without these measures, there's a risk of models overfitting to their training data, thus limiting their efficacy in diverse real-world scenarios.

\subsubsection{Challenges in LLM Interpretability, Trustworthiness, and Ethical Usage.} 
Ensuring interpretability and trustworthiness is paramount when integrating LLMs into security tasks, particularly given the sensitive nature of security requirements and the need for rigorous scrutiny of model outputs. The challenge lies in comprehending how these models make decisions, as the black-box nature of LLMs often impedes explanations for why or how specific outputs or recommendations are generated for security needs. While LLM self-explanations are often post-hoc rationalizations rather than reflections of internal workings~\cite{freiberger2025explainable,sarkar2024largelanguagemodelsexplain}, research explores methods to enhance their transparency and utility in security contexts. Techniques include promoting step-by-step reasoning during generation (e.g., Chain-of-Thought) which can improve output quality~\cite{kojima2023large,freiberger2025explainable}, grounding explanations in source data using RAG or citation mechanisms, visualizing attention weights as proxies~\cite{10541203}, defining structured evaluation criteria to constrain the LLM's reasoning process, and applying established XAI methods like LIME and SHAP to explain the behavior of models used within security workflows~\cite{lim2025explicateenhancingphishingdetection}. Furthermore, assessing the reliability of LLM judgments can be aided by estimating output uncertainty or checking for consistency across multiple generated responses~\cite{varshney2023stitchtimesavesnine}. A promising approach involves using LLMs specifically as 'explanation translators,' taking technical outputs from XAI tools (like feature importance scores from LIME/SHAP) and converting them into accessible, natural language descriptions tailored for end-users, as demonstrated in phishing detection~\cite{lim2025explicateenhancingphishingdetection}.

Recent research~\cite{bugrepair15,wang2023challenge7} has also underscored that artificial intelligence-generated content (AIGC) introduces additional security risks, including privacy breaches, dissemination of forged information, and the generation of vulnerable code. The absence of interpretability and trustworthiness can breed user uncertainty and reluctance, as stakeholders may hesitate to rely on LLMs for security tasks without a clear understanding of their decision-making process or adherence to security requirements. Establishing trust in LLMs necessitates further development and application of technologies and tools that offer deeper insights into model internals and output rationale, empowering developers and users to comprehend and verify the reasoning~\cite{freiberger2025explainable}. Improving interpretability and trustworthiness through such methods can ultimately foster the widespread adoption of cost-effective automation in the cybersecurity domain, fostering more efficient and effective security practices. However, many LLMs lack open-source availability, and questions persist regarding the data on which they were trained, as well as the quality, sources, and ownership of the training data, raising concerns about ownership regarding LLM-generated tasks. Moreover, there is the looming threat of various adversarial attacks, including tactics to guide LLMs to circumvent security measures and expose their original training data~\cite{challenge8}.

\subsubsection{Limitations in Modern Cryptanalysis}
A significant challenge to the broad applicability of LLMs is evident in the field of modern cryptanalysis. Current LLMs operate as probabilistic models, proficient at identifying statistical patterns within extensive textual datasets~\cite{gu2024survey}. This operational paradigm inherently conflicts with the deterministic, precise, and mathematically rigorous demands of cryptanalysis, which typically necessitates profound insights into areas like number theory and algebra to devise novel algorithms for computationally hard problems~\cite{swenson2008modern, bogdanov2016towards}. LLMs demonstrate notable limitations in executing complex multi-step mathematical computations, engaging in formal logical deduction, and managing abstract conceptualizations, often leading to errors or diminished performance when faced with irrelevant contextual information~\cite{ahn2024large, havrilla2024understanding}. Moreover, the data-centric learning approach of LLMs is fundamentally misaligned with the objective of discovering new attack methodologies for contemporary ciphers, primarily because datasets illustrating successful exploits against such robust systems are, by design, unavailable for training. The tendency of LLMs to generate specious yet plausible information (i.e., hallucinations) further compromises their utility in cryptanalytic applications where exactitude is non-negotiable \cite{zhang2023opp5}. Thus, the core architecture of LLMs as sophisticated language processors, rather than as deterministic logical inference engines, establishes a fundamental incompatibility with the stringent requirements for breaking modern cryptographic systems.

\subsection{Opportunities}
\subsubsection{Improvement of LLM4Security.} 
\
\newline\textbf{Training models for security tasks.} Deciding between commercially available pre-trained models like GPT-4~\cite{gpt4} and open-source frameworks such as T5~\cite{raffel2023T5} or LLaMa~\cite{touvron2023llama} presents a nuanced array of choices for tailoring tasks to individual or organizational needs. The distinction between these approaches lies in the level of control and customization they offer. Pre-trained models like GPT-4 are generally not intended for extensive retraining but allow for quick adaptation to specific tasks with limited data, thus reducing computational overhead. Conversely, frameworks like T5 offer an open-source platform for broader customization. While they undergo pre-training, researchers often modify the source code and retrain these models on their own large-scale datasets to meet specific task requirements~\cite{opp3}. This process demands substantial computational resources, resulting in higher resource allocation and costs, but provides the advantage of creating highly specialized models tailored to specific domains. Therefore, the main trade-off lies between the user-friendly nature and rapid deployment offered by models like GPT-4 and the extensive task customization capabilities and increased computational demands associated with open-source frameworks like T5.

\textbf{Inter-model interaction of LLMs.} Our examination indicates that LLMs have progressed significantly in tackling various security challenges. However, as security tasks become more complex, there's a need for more sophisticated and tailored solutions. As outlined in Section~\ref{rq3}, one promising avenue is collaborative model interaction through external augmentation methods. This approach involves integrating multiple LLMs~\cite{yang2023progfuzz6} or combining LLMs with specialized machine learning models~\cite{tang2023bugrepair14,cai2024harmful3} to improve task efficiency while simplifying complex steps. By harnessing the strengths of different models collectively, we anticipate that LLMs can deliver more precise and higher-quality outcomes for intricate security tasks.

\textbf{Impact and applications of ChatGPT.} In recent academic research, ChatGPT has garnered considerable attention, appearing in over half of the 185 papers we analyzed. It has been utilized to tackle specific security tasks, highlighting its growing influence and acceptance in academia. Researchers have favored ChatGPT due to its computational efficiency, versatility across tasks, and potential cost-effectiveness compared to other LLMs and LLM-based applications~\cite{laskar2023opp1}. Beyond generating task solutions, ChatGPT promotes collaboration, signaling a broader effort to integrate advanced natural language understanding into traditional cybersecurity practices~\cite{deng2023pentest1,qammar2023opp2}. By closely examining these trends, we can anticipate pathways for LLMs and applications like ChatGPT to contribute to more robust, efficient, and collaborative cybersecurity solutions. These insights highlight the transformative potential of LLMs in shaping the future cybersecurity landscape.

\subsubsection{Enhancing LLM’s Performance in Existing Security Tasks.} 
\
\newline\textbf{External retrieval and tools for LLM.} LLMs have demonstrated impressive performance across diverse security tasks, but they are not immune to inherent limitations, including a lack of domain expertise~\cite{kandpal2023opp6}, a tendency to generate hallucinations~\cite{zhang2023opp5}, weak mathematical capabilities, and a lack of interpretability. Therefore, a feasible approach to enhancing their capabilities is to enable them to interact with the external world, acquiring knowledge in various forms and manners to improve the factualness and rationality of generated security task solutions. One viable solution is to provide external knowledge bases for LLMs, augmenting content generation with retrieval-based methods to retrieve task-relevant data for LLM outputs~\cite{gao2024opp4,du2023bugde6}. Another approach is to incorporate external specialized tools to provide real-time interactive feedback to guide LLMs~\cite{alrashedy2024vulrepair6,armengolestapé2024reverse4}, combining the results of specialized analytical tools to steer LLMs towards robust and consistent security task solutions. We believe that incorporating external retrieval and tools is a competitive choice for improving the performance of LLM4Security.

\textbf{Addressing challenges in specific domains.} Numerous cybersecurity domains, such as network security and hardware security, encounter a dearth of open-source datasets, impeding the integration of LLMs into these specialized fields~\cite{sun2023contrast1}. Future endeavors may prioritize the development of domain-specific datasets and the refinement of LLMs to address the distinctive challenges and nuances within these domains. Collaborating with domain experts and practitioners is crucial for gathering relevant data, and fine-tuning LLMs with this data can improve their effectiveness and alignment with each domain's specific requirements. This collaborative approach helps LLMs address real-world challenges across different cybersecurity domains~\cite{bubeck2023opp7}.

\subsubsection{Expanding LLM’s Capabilities in More Security Domains.} 
\
\newline\textbf{Integrating new input formats.} In our research, we noticed that LLMs in security tasks typically use input formats from code-based and text-based datasets. The introduction of new input formats based on natural language, like voice and images, as well as multimodal inputs such as video demonstrations, presents an opportunity to enhance LLMs' ability to understand and process various user needs~\cite{yin2024opp8}. Integrating speech can improve user-model interaction, allowing for more natural and context-rich communication. Images can visually represent security task processes and requirements, providing LLMs with additional perspectives. Moreover, multimodal inputs combining text, audio, and visuals can offer a more comprehensive contextual understanding, leading to more accurate and contextually relevant security solutions.

\textbf{Expanding LLM applications.} We noticed that LLMs have received significant attention in the domain of software and system security. This domain undoubtedly benefits from the text and code parsing capabilities of LLMs, leading to tasks such as vulnerability detection, program fuzzing, and others. Currently, the applications of LLMs in domains such as hardware security and blockchain security remain relatively limited, and specific security tasks in certain domains have not yet been explored by researchers using LLMs. This presents an important opportunity: by extending the use of LLMs to these underdeveloped domains, we can potentially drive the development of automated security solutions.

\subsection{Roadmap}
We present a roadmap for future progress in utilizing Large Language Models for Security (LLM4Security), while also acknowledging the reciprocal relationship and growing exploration of Security for Large Language Models (Security4LLM) from a high-level perspective.

\textbf{Automating cybersecurity solutions.} The quest for security automation encompasses the automated analysis of specific security scenario samples, multi-scenario security situational awareness, system security optimization, and the development of intelligent, tailored support for security operatives, which possesses context awareness and adaptability to individual needs. Leveraging the generative prowess of LLMs can aid security operatives in comprehending requirements better and crafting cost-effective security solutions, thus expediting security response times. Utilizing the natural language processing capabilities of LLMs to build security-aware tools enables more intuitive and responsive interactions with security operatives. Moreover, assisting security operatives in fine-tuning LLMs for specific security tasks can augment their precision and efficiency, tailoring automated workflows to cater to the distinct demands of diverse projects and personnel.

\textbf{Incorporating security knowledge into LLMs.} A key direction for the future is to integrate specialized security task solutions and knowledge from the cybersecurity domain into LLMs to overcome potential hallucinations and errors~\cite{aghaei2023ids5,ling2023external1}. This integration aims to enhance LLMs' ability to address security tasks, especially those requiring a significant amount of domain expertise, such as penetration testing~\cite{deng2023pentest1,temara2023pentest3}, hardware vulnerability detection~\cite{10.5555/hardde0}, log analysis~\cite{liu2023log1,karlsen2023log3}, and more. Embedding rules and best practices from specific security domains into these models will better represent task requirements, enabling LLMs to generate robust and consistent security task solutions~\cite{cheng2025enhancingsemanticunderstandingpointer}.

\textbf{Security agent: integrating external augmentation and LLMs.} We have witnessed the unprecedented potential of applying LLMs to solve security tasks, almost overturning traditional security task solutions in LLM4Security~\cite{webfuzz2,chen2023vulde4,xia2024progfuzz4}. However, the inherent lack of domain-specific knowledge and hallucinations in LLMs restrict their ability to perceive task requirements or environments with high quality~\cite{zhang2023opp5}. AI Agents are artificial entities that perceive the environment, make decisions, and take actions. Currently, they are considered the most promising tool for achieving the pursuit of achieving or surpassing human-level intelligence in specific domains~\cite{xi2023road1}. We summarized the external enhancement techniques introduced in LLM4Security in Section~\ref{rq3}, optimizing LLMs' performance in security tasks across multiple dimensions, including input, model, and output~\cite{jin2023bugde2,perrina2023cti5,chen2023vulde4}. Security operators can specify specific external enhancement strategies for security tasks and integrate them with LLMs to achieve automated security AI agents with continuous interaction within the system.

\textbf{Multimodal LLMs for security.} In LLM4Security, all research inputs are based on textual language (text or code). With the rise of multimodal generative LLMs represented by models like Sora~\cite{road2}, we believe that future research in LLM4Security can expand to include multimodal inputs and outputs such as video, audio, and images to enhance LLMs' understanding and processing of security tasks. For example, when using LLMs as penetration testing tools, relevant images such as topology diagrams of the current network environment and screenshots of the current steps can be introduced as inputs. In addition, audio inputs (such as recordings of specific security incidents or discussions) can provide further background information for understanding security task requirements.

\textbf{Security for Large Language Models (Security4LLM).} LLMs have gained considerable traction in the security sector, showcasing their potential in security-related endeavors. Nonetheless, delving into the internal security assessment of LLMs themselves remains a pressing area for investigation~\cite{Yao_2024road3,pmlr-v235-huang24x,10.1145/3712001}. The intricate nature of LLMs renders them vulnerable to a wide range of attacks, necessitating innovative strategies to fortify the models~\cite{challenge8,liu2023road4,glukhov2023road7, Yao_2024road3,DBLP:conf/icse/ZhouL00SL025}. These vulnerabilities can be broadly categorized~\cite{Yao_2024road3}. AI-inherent vulnerabilities stem from the machine learning aspects and include adversarial attacks like data poisoning (manipulating training data to compromise the model~\cite{10.5555/3618408.3619882}) and backdoor attacks (embedding hidden triggers), as well as privacy-compromising inference attacks (attribute inference~\cite{staab2024memorizationviolatingprivacyinference}, membership inference~\cite{mireshghallah-etal-2022-empirical}) and extraction attacks (recovering sensitive training data~\cite{carlini2021extracting}). Instruction tuning attacks, such as jailbreaking (bypassing safety restrictions~\cite{challenge8}) and prompt injection (manipulating prompts to elicit unintended behavior~\cite{greshake2023youvesignedforcompromising}), exploit the way LLMs process instructions. Non-AI inherent vulnerabilities may arise from the surrounding infrastructure, such as remote code execution (RCE) in integrated applications~\cite{10.1145/3658644.3690338} or supply chain vulnerabilities involving third-party plugins~\cite{10.5555/3716662.3716715}.

Furthermore, LLMs can themselves be misused as tools for offensive purposes~\cite{Yao_2024road3}). Their advanced generation capabilities can be exploited to create sophisticated phishing emails~\cite{10466545}, generate malware~\cite{10188649}, spread misinformation~\cite{DBLP:conf/iclr/ChenS24}, or facilitate social engineering attacks~\cite{Falade_2023}. User-level attacks are particularly prevalent due to the human-like reasoning abilities of LLMs~\cite{Yao_2024road3}. Considering that the inputs and outputs for security tasks often involve security-sensitive data (such as system logs or vulnerability code in programs)~\cite{qi2023log4,vulrepair2}, the potential for misuse or data leakage poses significant cybersecurity risks. 

Mitigating these threats requires a multi-faceted approach~\cite{Yao_2024road3}. Defense strategies can be implemented at different stages: improving model architecture (e.g., considering capacity or sparsity~\cite{li2022largelanguagemodelsstrong}), during training (e.g., corpora cleaning~\cite{subramani-etal-2023-detecting}, adversarial training~\cite{pmlr-v97-wang19f}, safe instruction tuning~\cite{bianchi2024safetytunedllamaslessonsimproving}, applying differential privacy~\cite{li2022largelanguagemodelsstrong}), and during inference (e.g., instruction pre-processing~\cite{jain2023baselinedefensesadversarialattacks}, in-process malicious use detection~\cite{wang-etal-2023-rmlm}, and output post-processing like self-critique~\cite{kadavath2022languagemodelsmostlyknow}). An intriguing avenue for future research is to empower LLMs to autonomously detect and identify their own vulnerabilities. Specifically, efforts could focus on enabling LLMs to generate patches for their underlying code, thus bolstering their inherent security, rather than solely implementing program restrictions at the user interaction layer. Given this scenario, future research should adopt a balanced approach, striving to utilize LLMs for automating cost-effective completion of security tasks while simultaneously developing and evaluating robust techniques to safeguard the LLMs themselves. This dual focus is pivotal for fully harnessing the potential of LLMs in enhancing cybersecurity and ensuring compliance with cyber systems.

\section{Conclusion}\label{sec:conclusion}
The rapid integration of LLMs is creating a paradigm shift in cybersecurity, particularly by revolutionizing how security is integrated into the software engineering lifecycle. In this systematic literature review of 185 seminal papers, we have not only mapped the landscape of LLM4Security but also synthesized several key insights that highlight this transformation:

\begin{itemize}
    \item \textbf{A Paradigm Shift from Analysis to Generation and Repair:} A primary contribution of LLMs to software engineering is moving beyond traditional defect detection to automated generation and repair. Decoder-only models, in particular, are being leveraged to create and validate patches for software vulnerabilities and bugs, representing a significant leap towards self-healing software.
    
    \item \textbf{Solving Data Scarcity through Generative Augmentation:} LLMs offer a novel solution to the long-standing problem of data scarcity in security research. Our review found a prominent trend of using LLMs to generate high-quality synthetic data---from vulnerable code snippets to textual repair suggestions---thereby creating richer and more diverse datasets for training robust security tools.
    
    \item \textbf{The Rise of Hybrid Intelligence Workflows:} The most effective applications of LLMs do not operate in isolation. Instead, they act as intelligent "brains" orchestrating traditional software engineering tools like static analyzers, fuzzers, and verifiers. This hybrid, neuro-symbolic approach, which often involves external augmentation and tool integration, is becoming the de facto standard for building reliable LLM-powered security solutions.

    \item \textbf{The Emergence of Autonomous Agents:} Beyond single-task automation, a forward-looking trend is the development of autonomous LLM-based agents capable of handling complex, multi-step security workflows. These agents, which can plan, use tools, and learn from interactions, are being applied to entire vulnerability lifecycles and penetration testing, pointing towards a future of autonomous security operations.
\end{itemize}

These insights are the culmination of our in-depth analysis guided by four Research Questions (RQs), which systematically examined the security tasks (RQ1), LLM models (RQ2), adaptation techniques (RQ3), and data handling practices (RQ4) in the field. By also outlining the critical challenges and providing a roadmap for future work, this survey serves as a valuable resource for researchers and practitioners aiming to navigate and contribute to the rapidly evolving intersection of large language models and cybersecurity.

\bibliographystyle{ACM-Reference-Format}
\bibliography{reference}

\end{document}